\documentclass{IEEEtran}

\usepackage{subcaption}
\usepackage{amsmath,amsfonts,amssymb,amscd}
\usepackage{graphicx}
\usepackage{bm}
\usepackage{xcolor}
\usepackage{stfloats}
\usepackage[english]{babel}
\usepackage{textcomp}
\usepackage{algorithmic}
\usepackage{tcolorbox}
\usepackage{tabstackengine}
\setstackEOL{\cr}

\newcommand{\la}{\lambda}
\newcommand{\ta}{\theta}
\newcommand{\be}{\beta}
\newcommand{\ga}{\gamma}

\newcommand{\R}{\mathbb{R}}

\newcommand{\C}{\mathcal{C}}
\newcommand{\fot}{\frac{1}{2}}
\newcommand{\ep}{\epsilon}
\newcommand{\hp}{\hspace{-3mm}}

\newcommand{\om}{\omega}

\newcommand{\pa}{\partial}

\newcommand{\di}{{\rm d}}

\allowdisplaybreaks[4]

\newif\ifdraft
\drafttrue

\newtheorem{proposition}{\bfseries Proposition}
\newtheorem{example}{\bfseries Example}
\newtheorem{assumption}{\it Assumption}
\newtheorem{theorem}{\bfseries Theorem}
\newtheorem{corollary}{\bfseries Corollary}
\newtheorem{lemma}{\bfseries Lemma}

\newtheorem{remark}{\bfseries Remark}

\newtheorem{problem}{\bfseries Problem}

\def\BibTeX{{\rm B\kern-.05em{\sc i\kern-.025em b}\kern-.08em
    T\kern-.1667em\lower.7ex\hbox{E}\kern-.125emX}}

\title{\LARGE \bf
Proxy Control Barrier Functions: Integrating Barrier-Based and Lyapunov-Based Safety-Critical Control Design
}

\author{}

\author{Yujie Wang and Xiangru Xu\thanks{Y. Wang and X. Xu are with the Department of Mechanical Engineering, University of Wisconsin-Madison,
        Madison, WI 53706, USA. Email: 
\{yujie.wang, xiangru.xu\}@wisc.edu.}}

\begin{document}
\maketitle

\begin{abstract}
This work introduces a novel Proxy Control Barrier Function (PCBF) scheme that integrates barrier-based and Lyapunov-based safety-critical control strategies for strict-feedback systems with potentially unknown dynamics. The proposed method employs a modular design procedure, decomposing the original system into a proxy subsystem and a virtual tracking subsystem that are controlled by the control barrier function (CBF)-based and Lyapunov-based controllers, respectively. By integrating these separately designed controllers, the overall system's safety is ensured. Moreover, a new filter-based disturbance observer is utilized to design a PCBF-based safe controller for strict-feedback systems subject to mismatched disturbances. 
This approach broadens the class of systems to which CBF-based methods can be applied and significantly simplifies CBF construction by requiring only the model of the proxy subsystem. 
The effectiveness of the proposed method is demonstrated through numerical simulations. 
\end{abstract}

\section{Introduction}
\label{sec:introduction}
Control Barrier Functions (CBFs) have emerged as a powerful tool for designing controllers that ensure safety in the form of set invariance \cite{ames2016control,cohen2024safety,jankovic2018robust,krstic2023inverse,molnar2023safety,nguyen2021robust,tan2021high,taylor2022safe,wang2023safe}. {\color{black}When the reference trajectory is outside the safe region (e.g., the dynamic obstacle is unknown to the offline motion planning algorithm) or the nominal controller can lead to unsafe behaviors, CBFs can be employed as safety filters to alter control inputs in a minimally invasive manner.}
Compared with Lyapunov-based safe control methods such as Barrier Lyapunov Functions (BLFs) \cite{tee2009barrier} and Prescribed Performance Control (PPC) \cite{bechlioulis2008robust}, CBF-based methods offer several advantages, including less structural restrictions on constraints and a decoupled design of the control objective (via the nominal controller) and safety specification (via the CBF). 
However, despite many recent advances, there are certain systems that CBFs cannot handle but have been extensively studied by Lyapunov-based methods in the context of stabilizing control design, such as systems with unknown control directions. This limitation is partly because CBFs lack some favorable structural properties of Lyapunov functions, such as positive definiteness. Therefore, expanding the system class applicable to CBF-based safe control design methods warrants further investigation.

\begin{figure}[!t]
 \vskip -2mm 
 \centering
\includegraphics[width=0.47\textwidth]{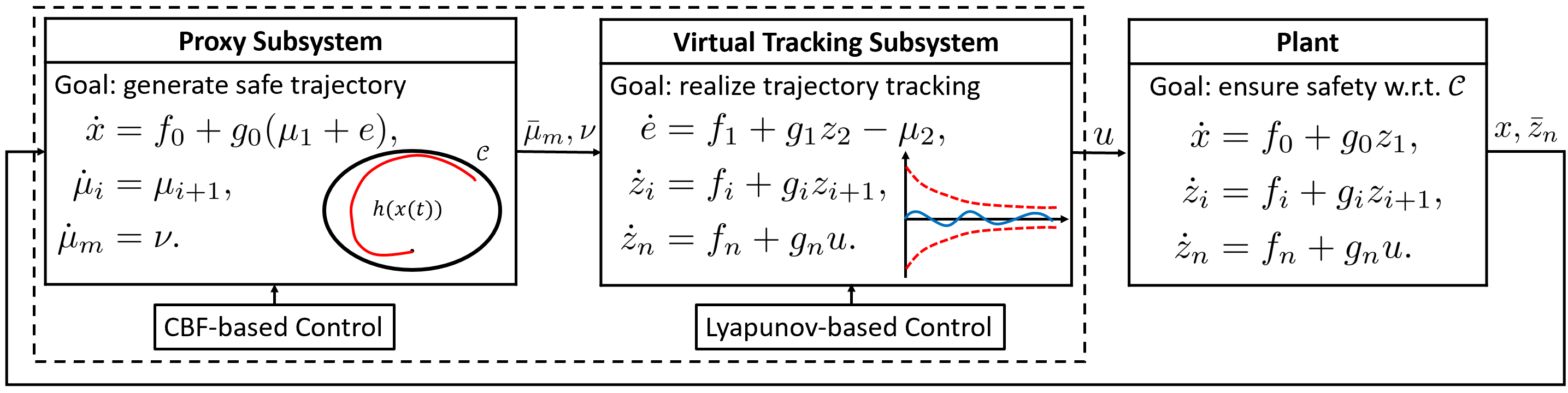}\vskip -2mm
\caption{Illustration of the proposed PCBF control scheme. The original system is decomposed into two subsystems that are controlled by CBF-based and Lyapunov-based methods, respectively. The PCBF method is modular  and inherits advantages of both CBF-based and Lyapunov-based approaches.
}
\label{fig:illustraion}
\end{figure}

Constructing \emph{valid} CBFs remains a challenging problem, especially for complex, high-dimensional systems or when disturbances and uncertainties are present. One promising solution is to construct CBFs based on a Reduced Order Model (ROM) of the full-order system, which  can significantly simplify the CBF construction and the safe control design process 
\cite{cohen2024safety,molnar2023safety,taylor2022safe}.
However, existing approaches either do not consider disturbances/uncertainties or rely on specific Lyapunov conditions (for tracking controllers) that are difficult to satisfy for certain systems.

This paper introduces a Proxy Control Barrier Function (PCBF) control strategy that follows a modular design scheme to integrate CBF-based and Lyapunov-based methods for strict-feedback systems with potentially unknown  dynamics. Specifically, the original system is decomposed into a proxy subsystem and a virtual tracking subsystem; the proxy subsystem generates a safe (virtual) reference trajectory and is controlled by a CBF-based controller, while the virtual tracking subsystem is controlled by a Lyapunov-based output-constrained controller to ensure the boundedness of the tracking error  (see Fig. \ref{fig:illustraion}). 
The modularity of the proposed PCBF method offers enhanced flexibility in control design, effectively combining the strengths of both CBF-based and Lyapunov-based approaches. By utilizing Lyapunov-based tools, the PCBF method extends applicability to systems that existing CBF-based methods cannot handle, particularly those with unknown dynamics. Furthermore, the CBF design process in the PCBF method is notably simplified, as the validity of the CBF does not depend on the full system dynamics, which allows systems with the same proxy subsystems to share a common CBF design. Compared to Lyapunov-based output-constrained control techniques (e.g., BLF and PPC), the PCBF method retains the advantages of CBF approaches, enabling a decoupled design of control objectives and safety specifications while accommodating more flexible safety constraint structures.

The contribution of this paper is twofold: (i) A PCBF control design scheme that integrates CBF-based and Lyapunov-based methods is proposed for strict-feedback systems with potentially unknown models; (ii) A PCBF-based control approach is developed for strict-feedback systems with mismatched disturbances using a new filter-based Disturbance Observer (DOB). 
A preliminary version of the paper appeared in \cite{wang2023safe} which only considered Euler-Lagrange systems. This paper extends \cite{wang2023safe} by studying the more general strict-feedback systems and the mismatched disturbance case.

The remainder of this paper is organized as follows: the motivation and problem formulation are given in Sec. \ref{sec:motivation}, the PCBF scheme is presented in Sec. \ref{sec:main}, the DOB-PCBF-based control strategy is presented in Sec. \ref{sec:dobpcbf}, and finally, the conclusion is drawn in Sec. \ref{sec:conclusion}.

\textit{Notation}: Given a positive integer $i$, $[i]=\{1,2,\cdots,i\}$. Given $z_i\in \R^{n_i}$ for $i\in[m]$, $\bar z_m=[z_1^\top z_2^\top \dots z_m^\top]^\top\in\R^{n_1+\cdots+n_m}$. 
Given $x\in\R^n$, $\|x\|$ represents its 2-norm. Given a function $f:\R\to \R$, $f^{(i)}$ represented its $i$-th derivative. For a square matrix $A$, $\lambda_\text{max}(A)$ and $\la_\text{min}(A)$ denote the maximal and minimal eigenvalues of $A$, respectively. Define $\R_{\geq 0}=\{x\in\R: x\geq 0\}$. Consider the gradient $\frac{\pa h}{\pa x}\in\R^{n\times 1}$ as a row vector, where $x\in\R^n$ and $h:\R^n\to\R$ is a function with respect to $x$. Denote $I_n$ as the identity matrix of size $n$.

\section{Motivation  and Problem Formulation}
\label{sec:motivation}
\subsection{Motivation}\label{subsec:motivation}

Consider a control affine system  $\dot x = f(x)+g(x) u$, where $x\in\R^n$ is the state, $u\in\R^m$ is the control input, and $f: \R^n\to\R^n$ and $g: \R^n\to\R^{n\times m}$ are known and locally Lipchitz continuous functions. Define a safe set 
$\mathcal{C} = \{ x \in \R^n : h(x) \geq 0\}$, where $h:\R^n\to\R$ is a sufficiently smooth function. The function $h$ is called a CBF of (input) relative degree 1 if   
$
 \sup_{u\in\R^m}  \left[ L_f h + L_{g} h u + \gamma h\right] \geq 0
$ holds
for all $x\in\R^n$, where $\gamma>0$ is a given positive constant, and $L_fh=\frac{\pa h}{\pa x}f$ and $L_gh=\frac{\pa h}{\pa x}g$ are Lie derivatives \cite{ames2016control}. When the CBF condition $L_f h + L_{g} h u + \gamma h \geq 0$ is incorporated into a Quadratic Program (QP), the resulting CBF-QP-based controller can formally ensure the safety (i.e., $h(x(t))\geq 0$ for any $t\geq 0$) of the closed-loop system. 

Partly because CBFs lack favorable structural properties of Lyapunov functions (e.g., positive definiteness), there are certain systems that CBFs cannot handle but have been extensively studied by Lyapunov-based methods in the context of stabilizing control design. Consider the following Norrbin  model for ship steering \cite{du2014adaptive}: 
\begin{IEEEeqnarray}{rCl}
\IEEEyesnumber \label{ship:both}
\IEEEyessubnumber \label{ship1}
\dot x_1 &=& x_2, \\
\IEEEyessubnumber \label{ship2}
\dot x_2 &=& bu+\theta^\top \varphi(x_2),
\end{IEEEeqnarray}
where $x_1=\psi\in \R$ is the yaw angle, $x_2=\dot \psi\in \R$ is the yaw rate, $u\in\R$ is the rudder angle as the control input, $\theta = [-\frac{1}{T} \ -\frac{\alpha}{T}]^\top$, $\varphi(x_2)=[x_2 \ x_2^3]^\top$, $b=\frac{K}{T}$, $K>0$ is the gain constant, $T$ is the time constant that can be either positive or negative,  and $\alpha$ is the Norrbin coefficient determined via a spiral test. If an uncontrolled ship (when $u=0$) exhibits straight-line stability (i.e., it moves along a straight path), then $T>0$; otherwise $T<0$ (see \cite{du2014adaptive} and \cite[Section 5.5]{fossen1999guidance} for more details).
Thus, $\ta$ and $b$ are considered as unknown parameters, and the sign of $b$ is unknown.

Suppose that the goal is to design a safe controller for system \eqref{ship:both} with respect to a given safe set $\C=\{x_1: h(x_1)\geq 0\}$ with $h$ a continuously differentiable function. To the best of our knowledge, CBF-based control strategies are not yet developed for systems with unknown control coefficients; if robust CBF methods (e.g., \cite{nguyen2021robust}) are applied, the resulting QP could be infeasible since the sign of $b$ is unknown.  On the other hand, the Nussbaum-gain-based adaptive control methods have been developed to stabilize systems with unknown control coefficients \cite{liu2017barrier}. Then, one natural question is: can we combine CBF-based safe control methods and Lyapunov-based stabilizing control methods, such that we can bring the best from both worlds to design a safe controller?

Observing the structure of  system \eqref{ship:both}, one may consider the system as two subsystems with $x_2$ as the input in \eqref{ship1}. To design a safe controller, one potential idea is to design a CBF-based control law for \eqref{ship1} to generate a safe reference trajectory for $x_2$ that ensures $h(x(t))\geq 0$ for any $t\geq 0$, and a Lyapunov-based output-constrained controller for \eqref{ship2} to ensure the boundedness of the tracking error. In this work, we will systematize this idea and propose a modular safe control design method for strict-feedback systems with theoretical guarantees.

\subsection{Problem Formulation}
\label{subsec:problem}

Consider a strict-feedback system described as follows:
\begin{subequations}\label{eqnsys:both} 
\begin{align}
\dot x&=f_0(x)+g_0(x)z_1,\label{eqnsys:subi}\\
\dot z_1&=f_1(x,z_1)+g_1(x,z_1)z_2,\label{eqnsys:sub2}\\
&\vdots\nonumber\\
\dot 
z_{n-1}&=f_{n-1}(x,z_1,\dots,z_{n-1})+g_{n-1}(x,z_1,\dots,z_{n-1})z_n,\label{eqnsys:sub3}\\
\dot z_n&=f_n(x,z_1,\dots,z_n)+g_n(x,z_1,\dots,z_n)u,\label{eqnsys:subn}
\end{align}
\end{subequations}
where $x\in\R^p$, $z_i\in\R^{p_i}$, $i\in[n]$, are state variables, $u\in\R^q$ is the control input, $f_0$ and $g_0$ are sufficiently smooth and known functions, and $f_i,g_i$, $i\in[n]$, are sufficiently smooth functions that are \emph{possibly unknown}. 
Define a safe set $\C$ as follows:
\begin{equation}
    \C=\{x\in\R^p: h(x)\geq 0\}\label{setc}
\end{equation}
where $h:\R^p\to\R$ is a sufficiently smooth function. 

The problem investigated in this paper is stated as follows. 
\begin{problem}\label{prob1}
Given the system shown in \eqref{eqnsys:both} and the safe set $\mathcal{C}$ defined in \eqref{setc}, design a controller  $u$ such that the closed-loop system is safe with respect to $\C$, i.e., $h(x(t))\geq 0,\forall t\geq 0$. 
\end{problem}

The main challenge in solving Problem \ref{prob1} lies in the presence of unknown functions  $f_i$ and $g_i$, $i\in[n]$. 
To the best of our knowledge, no safe control design method exists for system \eqref{eqnsys:both} if $f_i$ and $g_i$ are completely unknown. For the special case where $f_i$ or $g_i$ is the sum of a known function and an unknown disturbance, most existing methods tend to be conservative because the ``worst-case" of the disturbance is considered. The PCBF method proposed in this paper is based on a modular design scheme that not only can solve safe control design problems for a more general class of systems but also offer better control performance than existing methods. 
Furthermore, for many practical systems, it is reasonable to assume exact knowledge of $f_0$ and $g_0$ (e,g., for Euler-Lagrange systems,  $f_0=0$ and $g_0=1$). This assumption can be relaxed, which will be investigated  in our future work.

\section{PCBF-based Control Design}
\label{sec:main}
The PCBF method follows a \emph{modular} design scheme and  consists of three parts: (i) decompose system \eqref{eqnsys:both} into a proxy subsystem and a virtual tracking subsystem,  (ii) design a CBF-based controller for the proxy subsystem to generate a (virtual) safe trajectory, and (iii) design a Lyapunov-based tracking controller for the virtual tracking subsystem to ensure the boundedness of the tracking error. Because the CBF-based and Lyapunov-based control design are decoupled, the PCBF method offers flexibility for safe control design as will be shown below.

\subsection{System Decomposition}
\label{subsec:virtualdecomposition}
We decompose the strict-feedback  system shown in \eqref{eqnsys:both}  into the \emph{proxy subsystem}:
\begin{tcolorbox}[colback=white,top=0mm,bottom=0mm,right=0mm,left=2mm]
\begin{subequations}\label{proxy} 
\begin{align}
\dot x &=f_0(x)+g_0(x)\mu_1+g_0(x)e,\label{proxy1}\\
\dot \mu_1 &= \mu_{2},\label{vis2}\\
&\vdots\nonumber\\
\dot \mu_{m-1} &= \mu_{m},\label{vis3}\\
\dot\mu_m &=\nu,\label{visn}
\end{align}
\end{subequations}
\end{tcolorbox}
\noindent 
and the \emph{virtual tracking subsystem}:
\begin{tcolorbox}[colback=white,top=0mm,bottom=0mm,right=0mm,left=2mm]
\begin{subequations}\label{vts:both} 
\begin{align}
\dot e &= f_1(x,e+\mu_1)+g_1(x,e+\mu_1)z_2-\mu_2,\label{vts:sub2}\\
z_2&=f_2(x,e+\mu_1,z_2)+g_2(x,e+\mu_1,z_2)z_3,\\
&\vdots\nonumber\\
\dot z_{n-1}&=f_{n-1}(x,e+\mu_1,z_2,\dots,z_{n-1})\nonumber\\
&\qquad +g_{n-1}(x,e+\mu_1,z_2,\dots,z_{n-1})z_n,\\
\dot z_n&=f_n(x,e+\mu_1,z_2,\dots,z_n)\nonumber\\
&\qquad+g_n(x,e+\mu_1,z_2,\dots,z_n)u,\label{vts:subn}
\end{align}
\end{subequations}
\end{tcolorbox}
\noindent where $e=z_1-\mu_1$ is the virtual tracking error, $\nu\in\R^{p_1}$ is the virtual control input to be designed, and $\mu_1,\mu_2,\cdots,\mu_m\in\R^{p_1}$ are virtual states with the number  $m\leq n$ a positive integer to be determined. The initial conditions of the virtual states are selected as $\mu_1(0)=z_1(0)$ and $\mu_i(0)=0$ for $i=2,\cdots,m$.  

The proxy subsystem \eqref{proxy} consists of dynamics of $x$ and a chain of integrators, which  mainly serve to provide the explicit forms of the derivatives of $\mu_1$ as will be explained later; the virtual tracking subsystem \eqref{vts:both}  consists of error  dynamics of $e$, which is derived from \eqref{eqnsys:sub2} and \eqref{vis2}, and  equations of $z_3,z_4,\dots,z_n$. The state variables  $x,z_1,\cdots,z_n$ evolve identically in the original system  \eqref{eqnsys:both} and subsystems \eqref{proxy}-\eqref{vts:both}  when given the same control input $u$.

Given the system decomposition shown in \eqref{proxy}-\eqref{vts:both}, the safe control design problem for system \eqref{eqnsys:both}  will be solved by accomplishing the following two tasks: 
\begin{itemize}
\item Design a CBF-based control law $\nu$ 
for the subsystem \eqref{proxy} to ensure $h(x(t))\geq 0,\forall t\geq 0$,  
under the assumption 
\begin{equation}
\|e(t)\|\leq \rho(t),\;\;\forall t\geq 0,\label{definerho}
\end{equation}
where $\rho:\R_{\geq 0}\to\R_{> 0}$ is a predefined smooth bounded function 
whose derivatives up to  $n$-th order are bounded.
\item  Design a Lyapunov-based control law $u$ 
for the subsystem \eqref{vts:both} to ensure \eqref{definerho} holds. 
\end{itemize}

As will be explained  in the next three subsections, these two tasks can be accomplished separately, and the controllers constructed for the two subsystems together provide a solution to Problem \ref{prob1}.

\subsection{Proxy Subsystem Control Design}\label{sec:proxy}
The main difficulty of designing a CBF-based controller for the proxy subsystem \eqref{proxy} lies in the existence of $e$, 
which is considered as a mismatched disturbance to be rejected and has a relative degree lower than that of the virtual input $\nu$, implying it is difficult to completely decouple $e$ \cite{yang2013nonlinear}. 
To address this issue, we define a set of functions as follows:
\begin{align}
&b_i (\bar\mu_{i},\bar y_i,y_0,x,t) =\mathcal{M}_i(f_0+g_0\mu_1)-\frac{\|\mathcal{M}_ig_0\|^2}{2\be_i}-\frac{\be_i}{2}\rho^2
\nonumber\\
&\quad+\la_ib_{i-1}+\frac{\pa b_{i-1}}{\pa t}+\sum_{j=1}^{i-1}\frac{\pa  b_{i-1}}{\pa \mu_j}\mu_{j+1}, \ i\in[m],\label{hi}
\end{align}
where 
\begin{align}
\mathcal{M}_i= \frac{1}{\xi}\sum_{j=0}^{i-1}\frac{\pa b_{i-1}}{\pa y_j}\frac{\pa h}{\pa x}y_{j+1}+\frac{\pa b_{i-1}}{\pa x},\;i\in[m+1],\label{mi}
\end{align}
$\xi,\beta_i,\la_i$ for $i\in[m]$ are positive constants, $y_0=\chi(h/\xi)$, $y_i=\chi^{(i)}(h/\xi)$ for $i\in[m]$, and $b_0=y_0$. Here,  $\chi:\R\to\R$ is a $(m+1)$-th order differentiable function  satisfying $\chi(0)=0$, $\chi(\tau)=1$ for $\tau\geq 1$, and  $\frac{\di \chi}{\di\tau}>0$ for $\tau<1$ \cite{tan2021high}.

With these notations, the following theorem presents a CBF-based control design method for proxy subsystem \eqref{proxy} to ensure $h(x(t))\geq 0,\forall t\geq 0$, under condition \eqref{definerho}. The proof of this theorem  is given in Appendix \ref{proofthm1}.

\begin{theorem}\label{theorem:proxy}
Consider the proxy subsystem \eqref{proxy} and the safe set $\C$ defined in \eqref{setc}. Suppose that condition \eqref{definerho} holds,  
and there exist $\xi\!>\!0$, $\!\la_i\!>\!0$, $\!i\!\in\!\![m\!+\!1]$, and $\!\beta_i\!>\!0$, $\!i\!\in\![m]$, such that \\
(i) for any  $x\in\C$, $L_{g_0}h=0 \Rightarrow h\geq \xi$ holds;\\
(ii) for any $t\geq 0$,  $\sum_{j=2}^{m+1}\frac{\be_{j-1}}{2}\left(\frac{\di}{\di t}+\la_{j} \right)\circ \cdots \circ\left(\frac{\di}{\di t}+\la_{m+1} \right)\circ \rho(t)^2\leq \Pi_{j=1}^{m+1}\la_j$ holds;\\
(iii) $y_0(0)> 0$ and $b_i(\bar\mu_{i}(0),\bar y_i(0),y_0(0),x(0),0)> 0$ for $i\in[m]$ where $b_i$ is defined in \eqref{hi}.\\
Then, for any $x\in\C$ and $\mu_1,\cdots,\mu_m\in\R^{p_1}$, the set 
\begin{align}
K_{BF}= \left\{ {\mathfrak v}\in\R^{p_1}: \psi_{0}+\psi_{1} {\mathfrak v}\geq 0\right\}\label{KBF}
\end{align}
is non-empty, where
\begin{IEEEeqnarray}{rCl}
\IEEEyesnumber \label{psi01}
\psi_0 &=& \frac{\pa b_{m}}{\pa t} +\sum_{j=1}^{m-1}\frac{\pa b_{m}}{\pa \mu_j}\mu_{j+1}+\mathcal{M}_{m+1}(f_0+g_0\mu_1)\nonumber\\
\IEEEyessubnumber\label{psi01:0}
&&+\la_{m+1} b_{m}-\|\mathcal{M}_{m+1} g_0\|\rho,\\
\IEEEyessubnumber\label{psi01:1}
\psi_1&=& \frac{\pa b_{m}}{\pa \mu_m},
\end{IEEEeqnarray}
with $\mathcal{M}_{m+1}$ defined in \eqref{mi}. Moreover, any Lipschitz continuous controller $\nu\in K_{BF}$
ensures $h(x(t))\geq 0$, $\forall t\geq 0$.
\end{theorem}

\begin{remark}
The function $b_i$ in \eqref{hi} is specifically designed to ensure both the non-emptiness of $K_{BF}$ and $\dot b_{i-1}+\la_ib_{i-1}\geq b_i$ for $i\in[m]$. This guarantees the implication $b_{i}\geq0 \Rightarrow b_{i-1}\geq 0$  for $i\in[m]$; consequently, $\nu\in K_{BF}\Rightarrow b_m\geq 0 \Rightarrow b_0\geq 0 \Leftrightarrow h\geq 0$. The construction of $b_i$ is achieved by computing  $\dot b_{i-1}$ by using the chain rule and accounting for the worst-case scenario of the virtual tracking error  $e$. Detailed steps can be found in the proof. In practice, the explicit expression of $b_i$ can be derived using symbolic computation tools. For the special case where $L_{g_0}h\neq 0$ for any $x\in\C$, the conclusion of Theorem \ref{theorem:proxy} remain valid by dropping Condition (i) and (ii) and without using the function $\chi$ (i.e., replace functions $y_0$ and $y_i$ in \eqref{hi} with $h$ and $0$, respectively).  The details are omitted due to the page limitation.
\end{remark}

The safe control law $\nu$ in Theorem \ref{theorem:proxy} is obtained by solving the following convex CBF-QP:%
\begin{align}\label{cbfqp1}
\min_{\nu} \quad & \|\nu-\nu_{d}\|^2\\
\textrm{s.t.} \quad & \psi_0+\psi_1 \nu\geq 0 \nonumber
\end{align}
where $\psi_0,\psi_1$ are given in \eqref{psi01}, $\nu_{d}$ is any  nominal control law that is possibly unsafe. Note that this QP is always feasible by the non-emptiness of the set $K_{BF}$.
 Since $\nu$ serves  as the control input to the proxy subsystem rather than the original system \eqref{eqnsys:both}, the nominal controller $\nu_d$ is typically not provided directly. The following result offers a  method for designing $\nu_d$. 
The proof of this Corollary is given in Appendix \ref{proofcorollarytracking}.
\begin{corollary}\label{corollary:tracking}
Consider the proxy subsystem \eqref{proxy} and the safe set $\C$ defined in \eqref{setc}. Suppose that the right inverse of $g_0$ exists 
for any $x\in\C$,  condition \eqref{definerho} holds, and $x_d(t)$ is a reference trajectory that is $(m+1)$-th order differentiable. Then the control law $\nu$  given as $\nu= \alpha_{m+1}$, which corresponds to the nominal controller $\nu_d$ in \eqref{cbfqp1}, will ensure the tracking error $x-x_d$ is globally Uniformly Ultimately Bounded (UUB),
where $\alpha_{m+1}$ is defined recursively according to 
$\alpha_1=-g_0^{\dagger}(k_0\ep_0+f_0-\dot x_{d})-\frac{g_0^\top\ep_0}{2c_0}$, $\alpha_2=\frac{\pa\alpha_1}{\pa t}+\frac{\pa\alpha_1}{\pa x}(f_0+g_0\mu_1)-\frac{\ep_1}{2c_1} \big\|\frac{\pa\alpha_1}{\pa x}g_0 \big\|^2-g_0^\top\ep_0-k_1\ep_1$, $\alpha_i=\frac{\pa\alpha_{i-1}}{\pa t}+\frac{\pa\alpha_{i-1}}{\pa x}(f_0+g_0\mu_1)+\sum_{j=1}^{i-2}\frac{\pa\alpha_{i-1}}{\pa \mu_j}\mu_{j+1}-\ep_{i-2}-\frac{\ep_{i-1}}{2c_{i-1}} \big\|\frac{\pa\alpha_{i-1}}{\pa  x}g_0\big\|^2-k_{i-1}\ep_{i-1}$ for $i=3,\cdots,m+1$, with  $\ep_0=x-x_d$, $\ep_i=\mu_i-\alpha_i$ for $i\in[m]$, and  positive constants $k_i,c_i>0$ for  $i=0,1,\cdots,m$. 
\end{corollary}

\subsection{Virtual Tracking Subsystem Control Design}
\label{subsec:vts}

Control design for the virtual tracking subsystem \eqref{vts:both} can be accomplished by any Lyapunov-based method that ensures \eqref{definerho} holds, such as BLF \cite{jin2018adaptive,liu2017barrier,tee2009barrier} and PPC  \cite{bechlioulis2008robust,bechlioulis2014low}. This flexibility demonstrates modularity of our proposed approach.  

In particular, by leveraging the approximation-free PPC technique shown in \cite[Theorem 2]{bechlioulis2014low}, a ``model-free" control law without the information of $f_i$ and $g_i$  can be designed, as presented in the following result 
whose proof is similar to \cite[Theorem 2]{bechlioulis2014low} and omitted due to page limit. 
\begin{proposition}\label{theorem:modelfree}
Consider the virtual tracking subsystem \eqref{vts:both} with $q=1$ and $p_i=1$, $i\in [n]$. Suppose that (i) the sign of $g_i$ is known and $|g_i|\geq b_i$, $i\in[n]$, where $b_i>0$ is an unknown constant, and (ii) when $m>1$, $\mu_1$ and $\mu_2$  (or $\mu_1$ and $\nu$ when $m=1$) are bounded with possibly unknown bounds. Then, the control law designed as $u=\eta_n$ will ensure $|e(t)|\leq \rho(t)$ where $\eta_n$ is defined recursively according to $\eta_i=-k_i\log\left( \frac{1+\xi_i}{1-\xi_i}\right)$, $i\in[n]$, with $k_i$ a positive constant, $\xi_1=\frac{z_1-\mu_1}{\rho}$, $\xi_i=\frac{z_i-\eta_{i-1}(\bar z_{i-1},\mu_1,t)}{\rho_i}$ $(i=2,\cdots,n)$, and $\rho_i$ $(i=2,\cdots,n)$ smooth positive functions satisfying $\lim_{t\to\infty}\rho_i(t)>0$ and $\rho_i(0)>|z_i(0)-\eta_{i-1}(\bar z_{i-1}(0),\mu_1(0),0)|$.
\end{proposition}

The control law in Proposition \ref{theorem:modelfree} is robust since it does not rely on the information of $f_i$ and $g_i$, $i\in[n]$. Furthermore, 
because the reference signal in PPC is only required to be continuously differentiable \cite[Assumption 4]{bechlioulis2014low}, 
one may select $m = 1$ for the proxy subsystem, which will result in a simple control design in Subsection \ref{sec:proxy}. However, in this case, the PPC controller
tends to yield large and oscillating control input; see the simulation results in Example \ref{example:mismatch}. In addition, although we assume $p_i=1$ and $q=1$, the approximation-free PPC can be readily extended to multi-input multi-output systems (i.e., $p_i,q>1$), as discussed in \cite[Remark 2]{bechlioulis2014low}.

\subsection{Safety Guarantee of the Overall System}
\label{sec:overall}
The safety of system \eqref{eqnsys:both} can be ensured by combining the controllers separately designed for the proxy subsystem and the virtual tracking subsystem in the preceding subsections, as shown in the following result. 

\begin{corollary}\label{theorem:general}
Consider system \eqref{eqnsys:both}, the safe set defined in \eqref{setc}, and the decomposition shown in \eqref{proxy}-\eqref{vts:both}. Suppose that all conditions of Theorem \ref{theorem:proxy} are satisfied, such that a Lipschitz continuous controller $\nu\in K_{BF}$ is given by the CBF-QP \eqref{cbfqp1}. Then, any Lipschitz continuous control law $u$ that ensures $\|e(t)\|\leq \rho(t)$ for the virtual tracking subsystem \eqref{vts:both} will guarantee $h(x(t))\geq 0,\forall t\geq 0$, for system \eqref{eqnsys:both}. 
\end{corollary}

The modular safe control design method offers several advantages: i) By leveraging both CBF-based and Lyapunov-based tools, the proposed method can design safe controllers for a general class of systems shown in \eqref{eqnsys:both}  that cannot be tackled by either tool separately; ii) Because the proxy subsystem  \eqref{proxy} is more structured than the original system \eqref{eqnsys:both}, validity of the CBF $h$ can be verified by a simple condition (i.e., Condition (i) of Theorem \ref{theorem:proxy}), which significantly simplifies the CBF construction and guarantees feasibility of the CBF-QP shown in \eqref{cbfqp1} as $K_{BF}$ is non-empty;  
iii) Different systems (e.g., Euler-Lagrange systems) with the same proxy subsystem  share an  identical CBF design, which simplifies the whole safe control design process.

\begin{remark}\label{remark:direct}
Including a chain of integrators in the proxy subsystem is crucial. Suppose that the proxy subsystem is selected as $\dot x=f_0+g_0\nu+g_0e$ without integrators 
where $\nu$ is the virtual control and $e= z_1-\nu$. 
In this setup, the CBF design requires the constraint $\rho(0)>\|e(0)\|=\|z_1(0)-\nu(x(0),0)\|$ to be imposed on $\nu$, as  Lyapunov-based output-constrained control necessitates that the initial state remains within the output constraint.  Since $\rho(t)$ is used in designing $\nu(x,t)$, $\nu$ must first be constructed to ensure the safety of the proxy subsystem, followed by verification of whether the constraint $\rho(0)>\|e(0)\|$ is satisfied. This iterative process complicates the control design and may lead to scenarios where no control law simultaneously meets both the safety requirements and the  constraint $\rho(0)>\|e(0)\|$.
Moreover, the differentiability of $\nu$, which is required by most Lyapunov-based methods for the control design of the virtual tracking subsystem, is difficult to  guarantee as $\nu$ is the solution of a QP. 
Although this issue can be mitigated  by using smoothing techniques as part of the ROM-based method, 
such controllers tend to be more conservative than QP-based controllers and are challenging to construct when multiple safe constraints exist  \cite{cohen2023characterizing,ong2019universal}. In contrast, the proposed proxy subsystem formulation \eqref{proxy} can avoid these issues. 
\end{remark}

\begin{remark}\label{remark:romcomparison}
In contrast to the modular PCBF design method, the ROM-based methods involve coupled ROM safe control design and tracking control design \cite{cohen2024safety,taylor2022safe}. In these methods, the tracking controller must satisfy specific Lyapunov conditions (refer to \cite[Section 6.1]{cohen2024safety}), which can be challenging to achieve; for instance, the Lyapunov condition shown in \cite[eqn. (71)]{cohen2024safety} is not satisfiable by either the approximation-free PPC in Proposition \ref{theorem:modelfree} or the Nussbaum-based adaptive controller in Example \ref{example:ship}.
Furthermore, although the proxy subsystem includes additional integrators, verifying the validity of $h$ (i.e., Condition (i) in Theorem \ref{theorem:proxy}) is simpler compared to ROM-based methods (such as the equation above \cite[eqn. (38)]{cohen2024safety}) because $f_0$ is not involved.
\end{remark}
\begin{remark}
Although input constraints are not explicitly addressed in this work, they could be incorporated into the design process using techniques from input-constrained CBFs \cite{breeden2021high}, BLFs \cite{liu2019barrier}, and robust control \cite{wen2011robust}. Specifically, for proxy subsystems with polynomial dynamics,  sum-of-squares optimization offers a potential approach for designing CBFs that account for input constraints \cite{xu2017correctness}. We plan to explore this problem in future work.
\end{remark}

\begin{remark}
Similar to PPC \cite{bechlioulis2008robust}, the function $\rho(t)$ can be defined as an exponentially decaying function $\rho(t) = (\rho_0-\rho_\infty)e^{-\lambda_\rho t}+\rho_\infty$ 
where $\rho_0,\rho_\infty,\lambda_\rho$ are positive constants with $\rho_0>\rho_\infty$. In practice, choosing smaller $\rho_0$ and $\rho_\infty$, along with a larger $\la_\rho$, reduces the conservatism of
the CBF-QP controller; however, this may lead to larger control inputs for the PPC/BLF-based controllers.  Therefore, the choice of $\rho(t)$ should carefully balance the trade-off between improving control performance and limiting the magnitude of the control inputs.    
\end{remark}

\begin{example}\label{example:ship}
Consider system \eqref{ship:both} with parameters $T=31 , K= 0.5,\alpha=0.4$ as in \cite{du2014adaptive}. Recall that $\ta$ and $b$ are both unknown parameters, which makes the problem unsolvable by existing CBF methods. The safe set is given as $\C=\{\psi: \frac{\pi^2}{81}-\psi^2\geq 0\}$ that aims to keep $|\psi|\leq 20^\circ$;  the reference trajectory is  $x_{d}=30\sin(0.02t)$ in degrees.  
We select $\rho(t)=0.02$ and decompose the system into \eqref{proxy}-\eqref{vts:both} where the proxy subsystem is  $\dot x=\mu_1+e$, $\dot \mu_1=\nu$, and the virtual tracking subsystem is $\dot e=bu+\ta^\top \varphi-\nu$. With  parameters $\la_1=6,\la_2=1,\be_1=20,\xi=\frac{\pi^2}{81}$, one can verify that all conditions of Theorem \ref{theorem:proxy} are satisfied, so that $\nu$ is obtained by solving CBF-QP \eqref{cbfqp1}. We design a Nussbaum-based adaptive controller $u= N(\zeta)\alpha$ \cite{liu2017barrier}, where $\alpha=k e\! -\!\nu\! +\!\hat\theta^\top\varphi$, $N(\cdot)$ is a Nussbaum-type function, $\zeta$ and $\hat\theta$ are governed by adaptive laws $\dot \zeta =\frac{e \alpha}{\rho^2-e^2}$ and $\dot{\hat{\theta}}=\ga_1^{-1}\!\big(\!\frac{e\varphi }{\rho^2-e^2}\!\big)-\ga_2\hat\theta$, respectively, and $\ga_1=10,\ga_2=2,k=2$. By Corollary \ref{theorem:general}, the safety of the closed-loop system is satisfied. From the simulation result shown in Fig. \ref{fig:ship}, one can see that the trajectory of $\psi$ indeed stays within the safe region, while the desired tracking performance is well preserved inside $\C$.

\begin{figure}[!ht]
\centering
\includegraphics[width=0.235\textwidth]{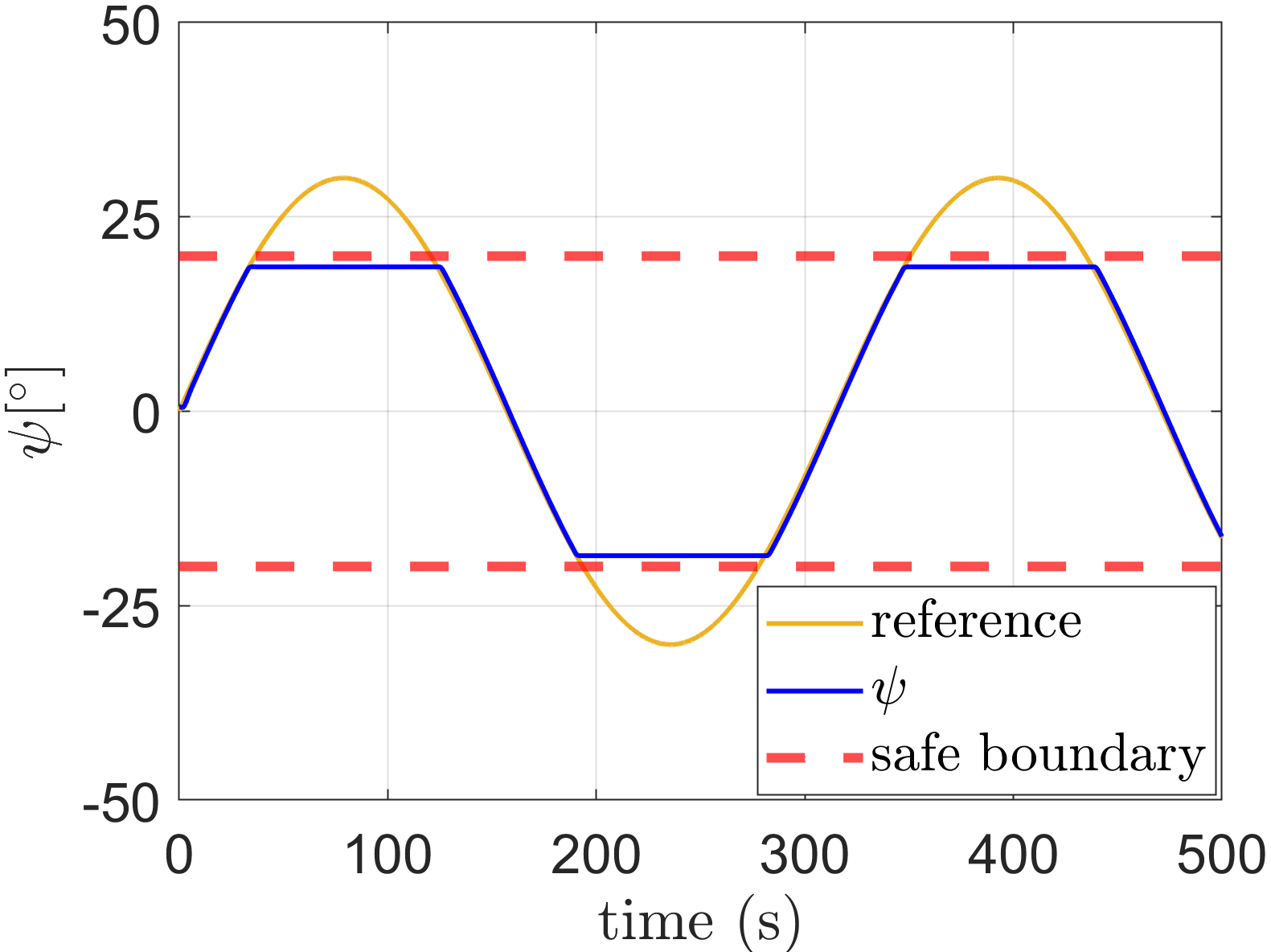}
\includegraphics[width=0.235\textwidth]{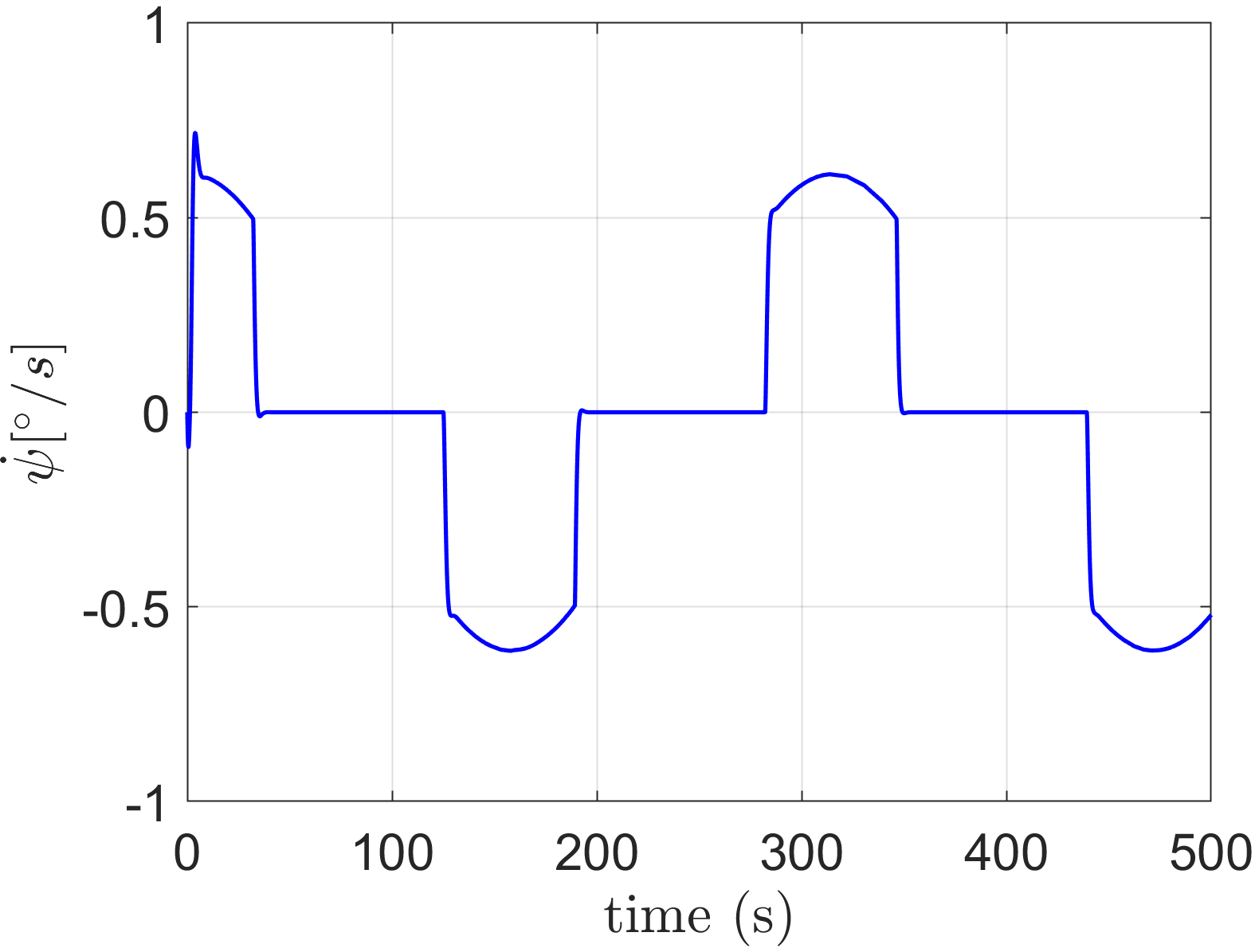}
\includegraphics[width=0.235\textwidth]{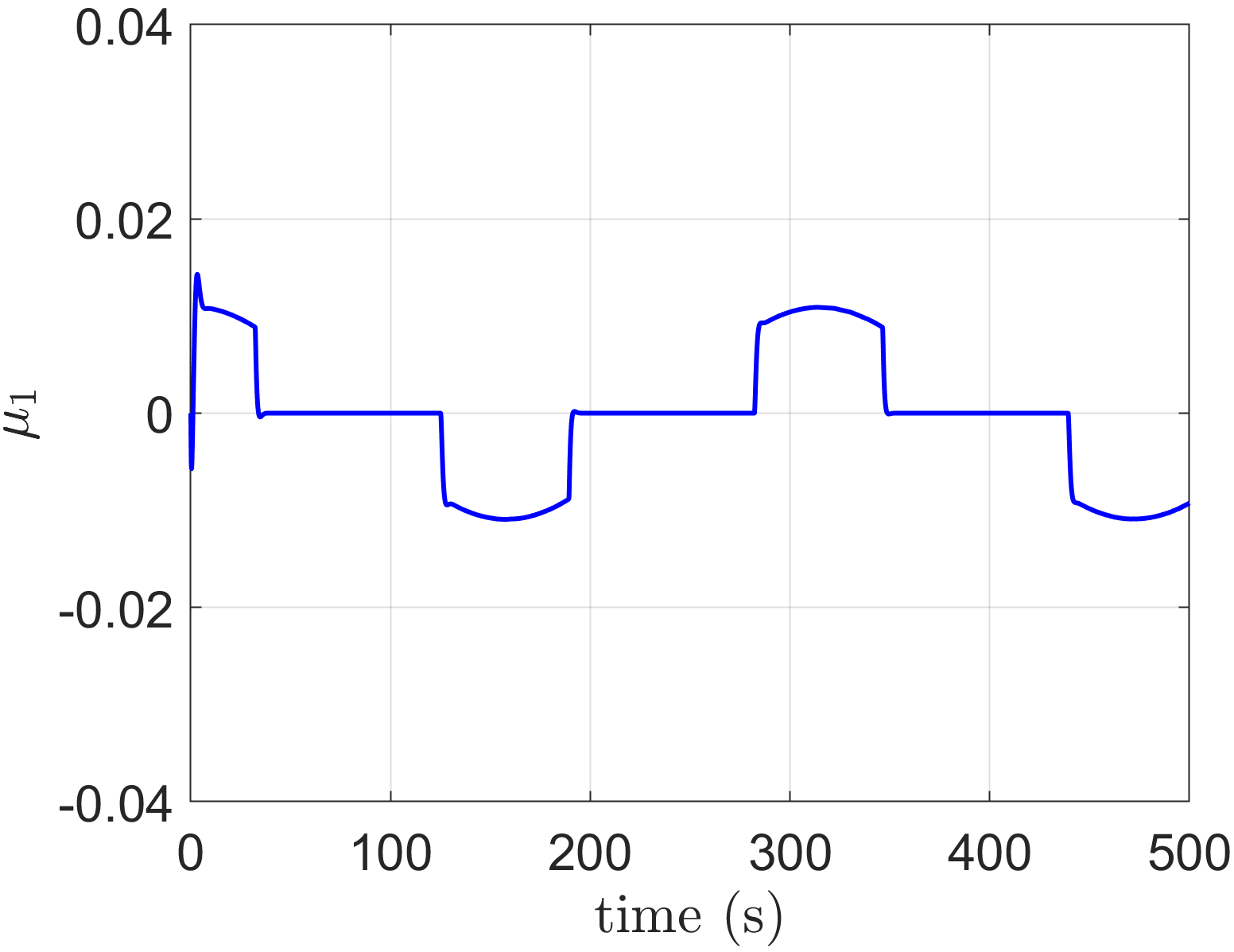}
\includegraphics[width=0.235\textwidth]{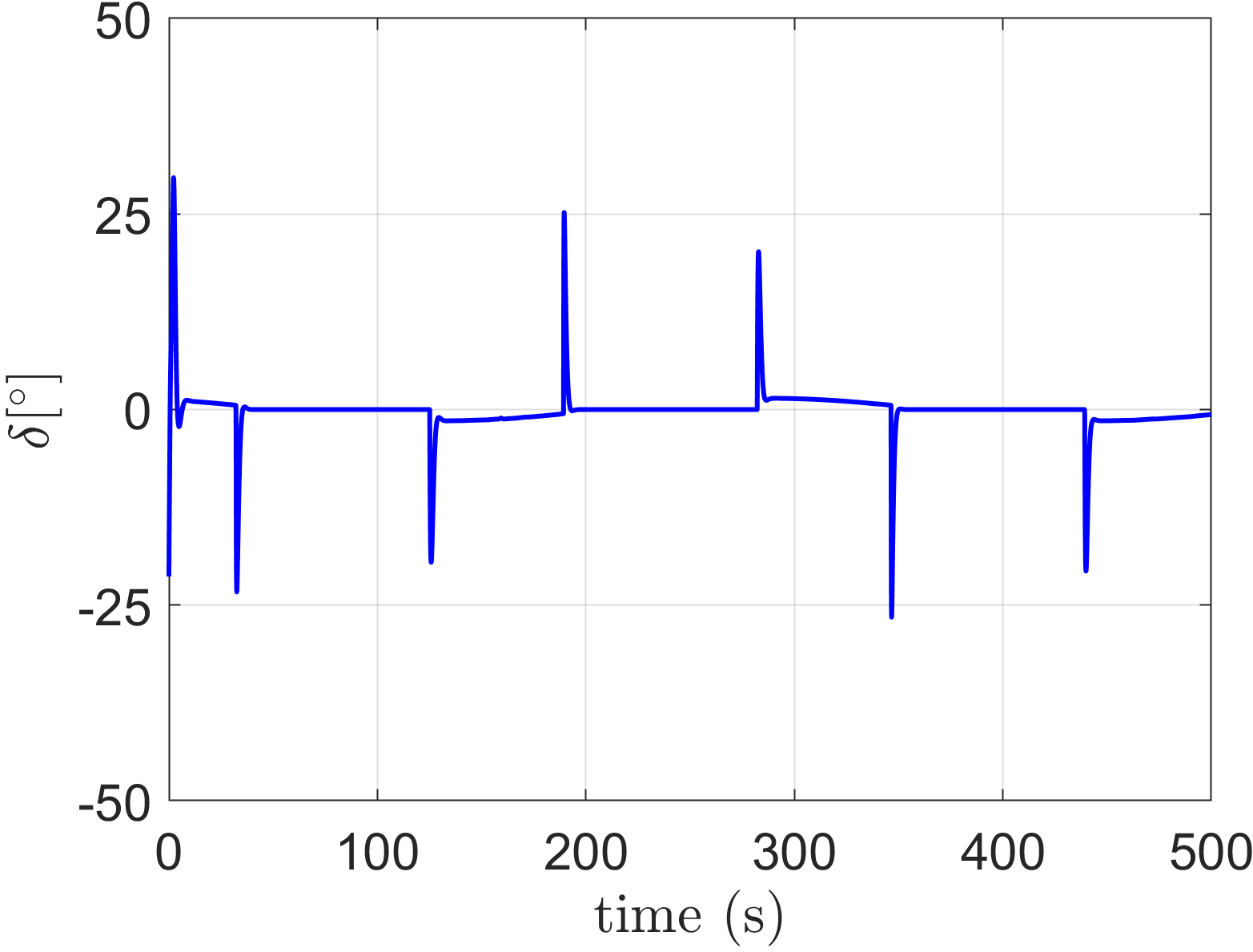}
  \caption{Simulation results of Example \ref{example:ship}. The safety is always respected as the trajectory of $\psi$ stays inside the safe region whose boundary is represented by the dashed red line.
  }
  \label{fig:ship}
\end{figure}
\end{example}

\section{DOB-PCBF Control Design}
\label{sec:dobpcbf}
In this section, we introduce a novel filter-based DOB and integrate it into the PCBF framework, enabling a DOB-PCBF-based safe control design scheme that addresses limitations of existing DOB-CBF methods in handling strict-feedback systems with mismatched disturbances.  The system under consideration is described as 
\begin{subequations}\label{systemmismatch}
\begin{align}
\dot x&=f_0(x)+g_0(x)z_1,\label{eqnsys:subidis}\\
\dot z_1&=f_1(x,z_1)+g_1(x,z_1)z_2+d_1,\label{eqnsys:sub2dis}\\
z_2&=f_2(x,z_1,z_2)+g_2(x,z_1,z_2)z_3+d_2,\label{eqnsys:sub3dis}\\
&\vdots\nonumber\\
\dot z_n&=f_n(x,z_1,\dots,z_n)+g_n(x,z_1,\dots,z_n)u+d_n,\label{eqnsys:subndis}
\end{align}
\end{subequations}
where $x\in\R^p,z_1\in\R^{p_1},\dots,z_n\in\R^{p_n}$ are state variables, $f_0,f_1,\dots,f_n$ and $g_0,g_1,\dots,g_n$ are all known sufficiently smooth functions, $d_1,\dots,d_n$ represent lumped unknown disturbances/uncertainties, and $u\in\R^q$ is the control input. We aim to design a controller  $u$ for system \eqref{systemmismatch} such that the closed-loop system is safe with respect to the set $\mathcal{C}$ defined in \eqref{setc}, i.e., $h(x(t))\geq 0,\forall t\geq 0$. Note that $f_i$ and $g_i$, $i\in [n]$ in system \eqref{systemmismatch} can be considered as the known part of the model with $d_i$ as the unknown part - this is different from the ``model-free'' problem setting in Proposition \ref{theorem:modelfree} where functions $f_i$ and $g_i$, $i\in [n]$, in system \eqref{eqnsys:both} are assumed to be  unknown. 

Robust CBF methods that consider the ``worst-case" of disturbances have been developed for disturbed systems, but their performance tends to be conservative \cite{jankovic2018robust,nguyen2021robust}.  To mitigate their unnecessary conservativeness, several DOB-CBF-based control schemes that precisely estimate and compensate for disturbances are proposed \cite{das2024robust,wang2022disturbance,wang2024immersion}. However, for systems whose disturbance relative degree is lower than the input relative degree (e.g., system \eqref{systemmismatch}), results on DOB-CBF-based control are still limited.  
For example, the method in \cite{wang2022disturbance} is overly conservative when the system dimension is greater than 2, and the method in \cite{das2024robust} is not applicable to systems whose input and disturbance relative degrees differ more than one.

To design a safe controller for \eqref{systemmismatch}, we will propose a new filter-based DOB and the corresponding  DOB-PCBF-based safe control strategy. The following assumption is standard.

\begin{assumption}\label{assumption:dob}
    The derivatives of disturbances $d_i$, $i\in[n]$, are bounded, i.e., $\|\dot d_i\|\leq\omega_i$, where $\omega_i$ is an unknown constant.
\end{assumption}

The DOB we use has the following form \cite{chen2015disturbance}:
\begin{subequations}\label{dob}
\begin{align}
\hat d_i&=s_i+ \alpha_i z_i,\\
\dot s_i&=\begin{cases}
-\alpha_i(f_i+g_i z_{i+1}+\hat d_i), & \text{if}\  i\in[n-1],\\
-\alpha_n(f_n+g_n u +\hat d_n), &\text{if}\ i=n,
\end{cases}
    \end{align}
\end{subequations}
where $\hat d_i$ is the estimation of $d_i$, $\alpha_i$ is a positive constant, and $s_i$ is the internal state of the DOB, for $i\in[n]$. With a Lyapunov function $V^d_i=\fot e_{d,i}^\top e_{d,i}$, 
where $e_{d,i}= \hat d_i-d_i$, one can verify that $\dot V^d_i\leq-2\kappa_iV^d_i+\omega_i^d$
where $\kappa_i=\alpha_i-\frac{\nu_i}{2}$, $\nu_i<2\alpha_i$ is a positive constant, and $\om_i^d=\frac{\om_i^2}{2\nu_i}$. Thus, $e_{d,i}$ is globally UUB.

Following the system decomposition scheme developed in Section \ref{sec:main}, system \eqref{systemmismatch} can be decomposed into the proxy subsystem shown in \eqref{proxy} with $m=n$  and the following virtual tracking subsystem:
\begin{IEEEeqnarray}{rCl}
    \IEEEyesnumber \label{systemmismatchvts}
    \IEEEyessubnumber\label{systemmismatchvts:1}
    \dot e&=& f_1+g_1 z_2+d_1-\mu_2,\\
    \IEEEyessubnumber\label{systemmismatchvts:2}
    \dot z_2&=& f_2+g_2 z_3+d_2,\\
    &\vdots&\nonumber\\
    \IEEEyessubnumber\label{systemmismatchvts:i}
    \dot z_n&=& f_n+g_n u+d_n,
\end{IEEEeqnarray}
where $e = z_1-\mu_1$ and $\mu_i$, $i\in[m]$, are defined in \eqref{proxy}.
The CBF-based control design for the proxy subsystem follows Theorem \ref{theorem:proxy} under condition \eqref{definerho}. We will propose a new DOB-based control approach for the subsystem \eqref{systemmismatchvts} to ensure \eqref{definerho} holds.

The high-order derivatives of $\hat d_i$, $i\in[n-1]$, are indispensable for control design because $d_i$ is a mismatched disturbance. However, $\dot{\hat{d}}_i$ is unknown because $\dot{\hat{d}}_i=-\alpha_i e_{d,i}$ and $e_{d,i}$ relies on $d_i$. To address this issue, we propose a filter-based DOB, which can generate alternative disturbance estimation signals that are close to $d_i$ and have known derivatives, as follows:
\begin{equation}
    \dot{\hat{d}}^f_{i,j}=-T_{i,j}(\hat d^f_{i,j}-\hat d^f_{i,j-1}) \label{lowpassfilter}
\end{equation} 
where $j\in[n-i], i\in[n-1]$, $\hat d_{i,0}^f=\hat d_i$, $\hat d_{i,j}^f$ is the filtered disturbance estimation, and $T_{i,j}>0$ is a positive constant. 
The following lemma shows the convergence of the filter. 
\begin{lemma}\label{corollary:filter}
Consider the DOB given in \eqref{dob} and the filter presented in \eqref{lowpassfilter}. Suppose that Assumption \ref{assumption:dob} holds. Then the filtered disturbance estimation error $\delta_i=\hat d^f_{i,n-i}- d_i$, $i\in[n-1]$, is globally UUB.
\end{lemma}

The proof of the lemma is given in Appendix \ref{proof:corollary3}. 
From Lemma \ref{corollary:filter} one can see that $\hat d_{i,n-i}^f$, whose derivatives up to the $(n-i)$-th order are explicitly given, is close to $d_i$ in the sense that $\|\delta_i\|$ is bounded by a known decaying function, implying that one can replace $\hat d_i$ with $\hat d_{i,j}^f$ in DOB design. Note that the ultimate bound of $\delta_i$ can be made arbitrarily small by selecting appropriate parameters.

The following theorem provides a filtered-DOB-based controller for the virtual tracking subsystem \eqref{systemmismatchvts} to ensure $\|e\|\leq\rho$. {\color{black} The control design follows backstepping \cite{KKK95}, and the filtered disturbance estimation error $\delta_i$ is compensated by virtual control signals. }
\begin{theorem}\label{theorem:systemmismatchblf}
Consider system \eqref{systemmismatch}, the safe set defined in \eqref{setc}, and the  decomposition given by \eqref{proxy} and \eqref{systemmismatchvts} with $m=n$. Suppose that all conditions of Theorem \ref{theorem:proxy} are satisfied, such that a Lipschitz continuous controller $\nu$ is given by the CBF-QP \eqref{cbfqp1}. Suppose that Assumption \ref{assumption:dob} holds, the right inverse of $g_i$ exists for any $x\in\C$ and $z_1\in\R^{p_1},\cdots,z_i\in\R^{p_i}$ $(i\in[n])$, the DOB is given in \eqref{dob}, and the filter is given in \eqref{lowpassfilter}. If  
    \begin{IEEEeqnarray}{rCl}
    u\!&=&\!g_n^{\dagger}\!\bigg[\!\sum_{j=1}^{n-1}\!\frac{\pa\tau_{n-1}}{\pa z_j}\hat d_{j,n\!-\!j}^f\!-\!\sum_{j=1}^{n-1}\!\frac{(n\!-\!j\!+\!1)\ep_n}{4\ga_j^f}\!\left\|\!\frac{\pa\tau_{n\!-\!1}}{\pa z_j}\!\right\|^2\!\!\!-\!\hat{d}_n\!-\!f_n\nonumber\\
\label{dobcontrol}
    &&-\frac{\ep_n}{4(\kappa_n-\sigma_n)}-k_{n}\ep_{n}\!-\!g_{n-1}^\top l_{n}\!+\!\mathcal{N}_{n}\!\bigg],
    \end{IEEEeqnarray}
    where
    \begin{IEEEeqnarray}{rCl}
    \IEEEyesnumber\label{systemmismatchcontroller}
    \IEEEyessubnumber\label{systemmismatchcontroller:1}
        \hspace{-9mm}\tau_{1}\!&=& \!g_1^{\dagger}\!\!\left[\!\frac{\dot\rho}{\rho}e\!-\!k_1e\!-\!\hat d_{1,n\!-\!1}^f\!\!-\!\frac{ne}{4(\!\ga_1^f\!\!-\!\sigma_1\!)(\!\rho^2\!-\!\|e\|^2\!)}\!-\!f_1\!+\!\mu_2\!\right]\!\!,\\
    \hspace{-9mm}\tau_{i}\!&=&\!g_i^{\dagger}\!\bigg[\!\sum_{j=1}^{i-1}\!\frac{\pa\tau_{i-1}}{\pa z_j}\hat d_{j,n\!-\!j}^f\!-\!\sum_{j=1}^{i-1}\!\frac{(n\!-\!j\!+\!1)\ep_i}{4\ga_j^f}\!\left\|\!\frac{\pa\tau_{i\!-\!1}}{\pa z_j}\!\right\|^2\!\!\!-\!\hat{d}_{i,n\!-\!i}^f\!-\!f_i\nonumber\\
\IEEEyessubnumber\label{systemmismatchcontroller:i}
    &&\!-\frac{(n\!-\!i\!+\!1)\ep_i}{4(\ga_i^f\!-\!\sigma_i)}\!-\!k_{i}\ep_{i}\!-\!g_{i-1}^\top l_{i}\!+\!\mathcal{N}_{i}\!\bigg], i\!=\!2,\cdots,n\!-\!1,
    \end{IEEEeqnarray}
{\color{black} are virtual control signals designed by following backstepping},
$\kappa_n$ is defined after \eqref{dob}, $k_i>0$ $(i\in[n])$, $\ga_i^f>0$ $(i\in[n-1])$,  $0<\sigma_i<\ga_i^f$ $(i\in[n-1])$, $0<\sigma_n<\kappa_n$, $\ep_1=e$, $\ep_i=z_i-\tau_{i-1}$ $(i=2,\cdots,n)$, $l_2=\frac{e}{\rho^2-\|e\|^2}$, $l_i=\ep_{i-1}$ $(i=3,\cdots,n)$, $\mathcal{N}_i=\frac{\pa\tau_{i-1}}{\pa t}+\sum_{j=1}^{i-1}\frac{\pa\tau_{i-1}}{\pa z_j}(f_j+g_jz_{j+1})+\sum_{j=1}^{i}\frac{\pa\tau_{i-1}}{\pa\mu_j}\mu_{j+1}-\sum_{j=1}^{i-1}\sum_{m=j}^{i-1}\frac{\pa\tau_{i-1}}{\pa \hat{d}_{j,n-m}^f}T_{j,n-m}(\hat d_{j,n-m}^f-\hat d_{j,n-m-1}^f)+\frac{\pa \tau_{i-1}}{\pa x}(f_0+g_0z_1)$ for $i=2,\cdots,n$, and $\mu_{n+1}=\nu$, then  \eqref{definerho} holds.
\end{theorem}

The proof of this theorem is given in Appendix \ref{proof:theorem2}. Safety of the closed-loop system \eqref{systemmismatch} with the controller $u$ follows from Corollary \ref{theorem:general} because of modularity of the proposed method.

\begin{remark}
The DOB-CBF-QP controllers developed in \cite{das2024robust,wang2022disturbance} do not compensate for all uncertain terms, and their performance heavily depends on the availability of accurate bounds for the disturbances or their derivatives, which are often difficult to determine in practice. While choosing sufficiently large bounds can ensure safety, it may lead to overly conservative control performance. In contrast, the method presented in Theorem \ref{theorem:systemmismatchblf} compensates for all unknown terms using the filters  and  does not rely on the bounds of the disturbances or their derivatives - this highlights the effectiveness of the modular PCBF control framework.
\end{remark}

\begin{example}\label{example:mismatch}
Consider the following electromechanical system  with mismatched disturbances adopted from \cite{yu2018finite}:
\begin{IEEEeqnarray}{rCl}
\IEEEyesnumber\label{electro}
\IEEEyessubnumber 
\dot x &=& z_1,\\
\IEEEyessubnumber
\dot z_1 &=& f_1(x,z_1)+g_1 z_2+d_1,\\
\IEEEyessubnumber 
\dot z_2 &=& f_2(x,z_1,z_2)+g_2u +d_2,
\end{IEEEeqnarray}
where $f_1 = -(Bz_1+N_0\sin(x))/M_0$, $g_1=1/M_0$, $f_2 = -(K_bz_1+R_mz_2)/L_m$, $g_2=1/L_m$, $d_1,d_2$ are external disturbances, and $M_0,B,N_0,L_m,R_m,K_b$ are physical parameters whose values are the same as those in \cite{yu2018finite}.
The safe set is $\mathcal{C} = \{ x \in \R^n : -0.5\leq x\leq 0.3 \}$, so we select two CBFs $h_1=x+0.5$ and $h_2=-x+0.3$. For comparison, we implement three methods in simulation: (i) the DOB-PCBF method by utilizing Corollary \ref{corollary:tracking}, Corollary \ref{theorem:general},  and Theorem \ref{theorem:systemmismatchblf}; (ii) the model-free PCBF-based control law by using Corollary \ref{corollary:tracking}, Corollary \ref{theorem:general}, and  Proposition \ref{theorem:modelfree}; and (iii) the DOB-CBF-QP proposed in \cite{wang2022disturbance}. The disturbances are selected as $d_1=d_2=\sin(t)+0.2\sin(2t)-0.5\cos(5t)+\cos(3t)$, from which one can verify that Assumption \ref{assumption:dob} holds with $\om_1=\om_2=6$; the reference trajectory is $x_{d}=\sin(t)$; and the control parameters for the DOB-PCBF method are $\be_1=\be_2=0.05,\rho(t)=0.8e^{-10t}+0.05, \la_1=\la_2=10,\la_3=15, \alpha_1=\alpha_2=30, \ga_1=50, \theta_1=10, T_{1,1}=100,k_1=k_2=10,\sigma_1=\sigma_2=15,\nu_1=\nu_2=1$. 
The Symbolic Math Toolbox in MATLAB is used to compute $\psi_0,\psi_1$ in \eqref{psi01}, as well as $u$ in \eqref{dobcontrol} and \eqref{systemmismatchcontroller}.
The simulation results are presented in Fig. \ref{fig:electromechanical}, which demonstrates that all three methods can ensure safety but their tracking performance and control input profiles are different: the DOB-PCBF method has perfect tracking performance with the smoothest control input profile; the model-free PCBF approach generates almost identical tracking performance as the DOB-PCBF method but its control input profile is more oscillatory; the DOB-CBF-QP approach has a poor tracking performance compared with other two methods.

\begin{figure}[!ht]
\centering
\includegraphics[width=0.23\textwidth]{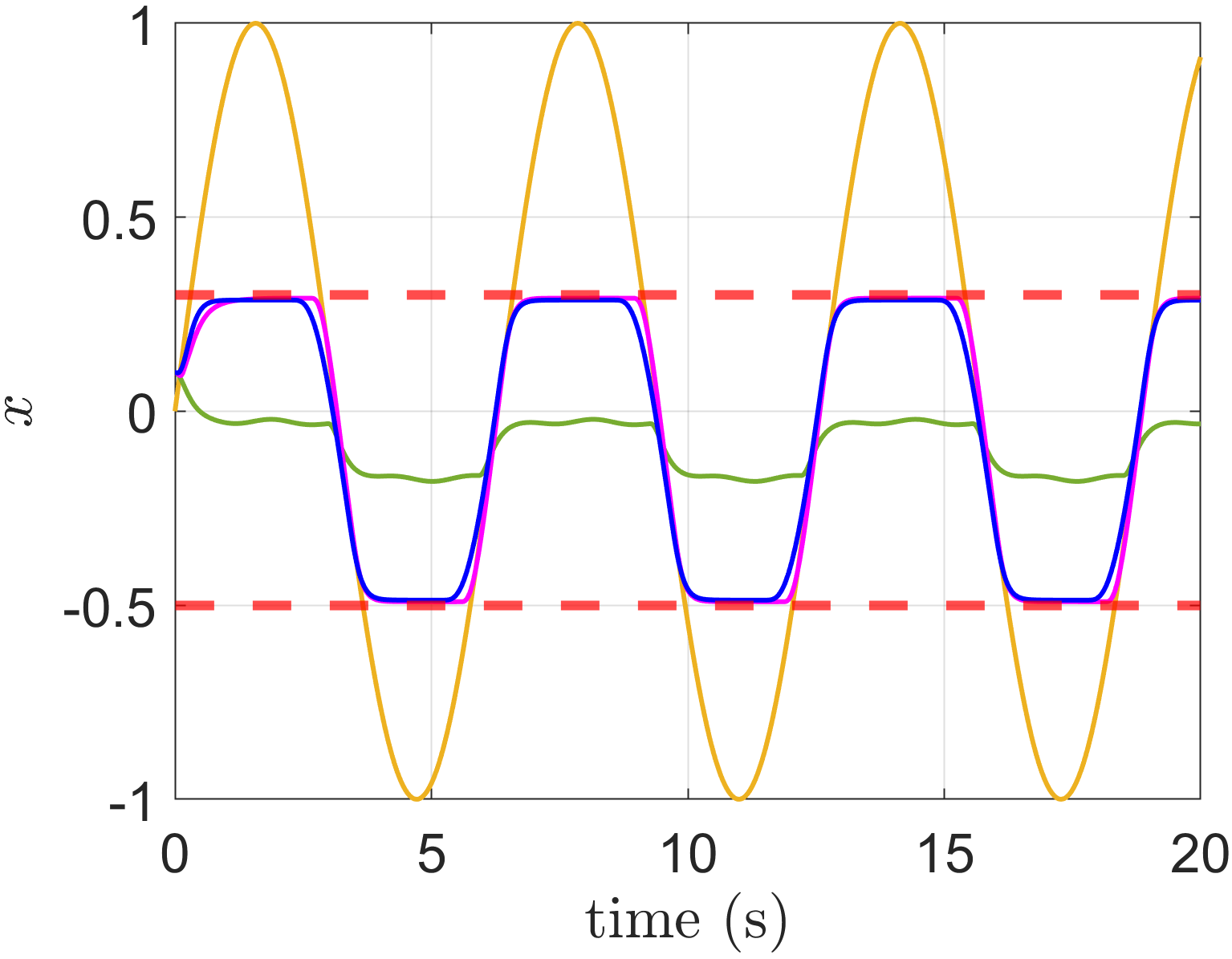}
\includegraphics[width=0.23\textwidth]{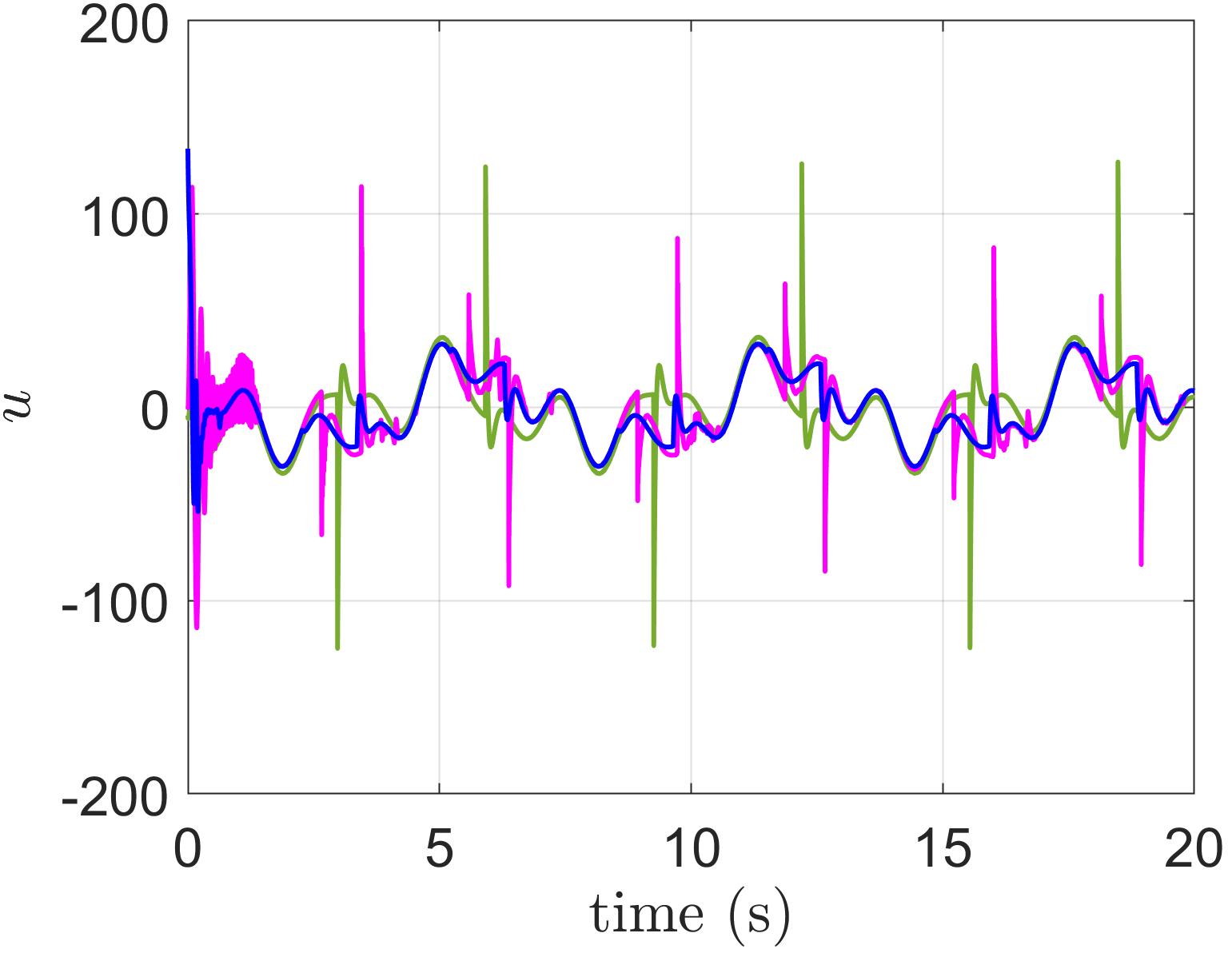}\vskip 2mm
\includegraphics[width=0.38\textwidth]{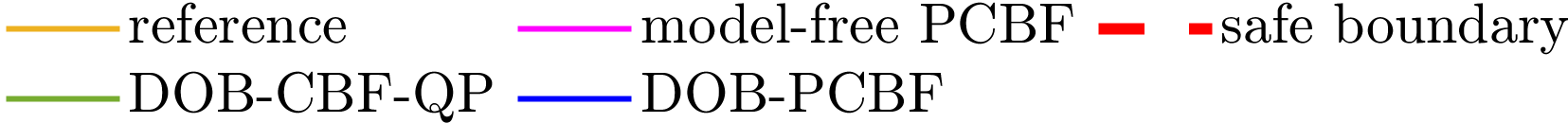}\vskip 2mm
\includegraphics[width=0.23\textwidth]{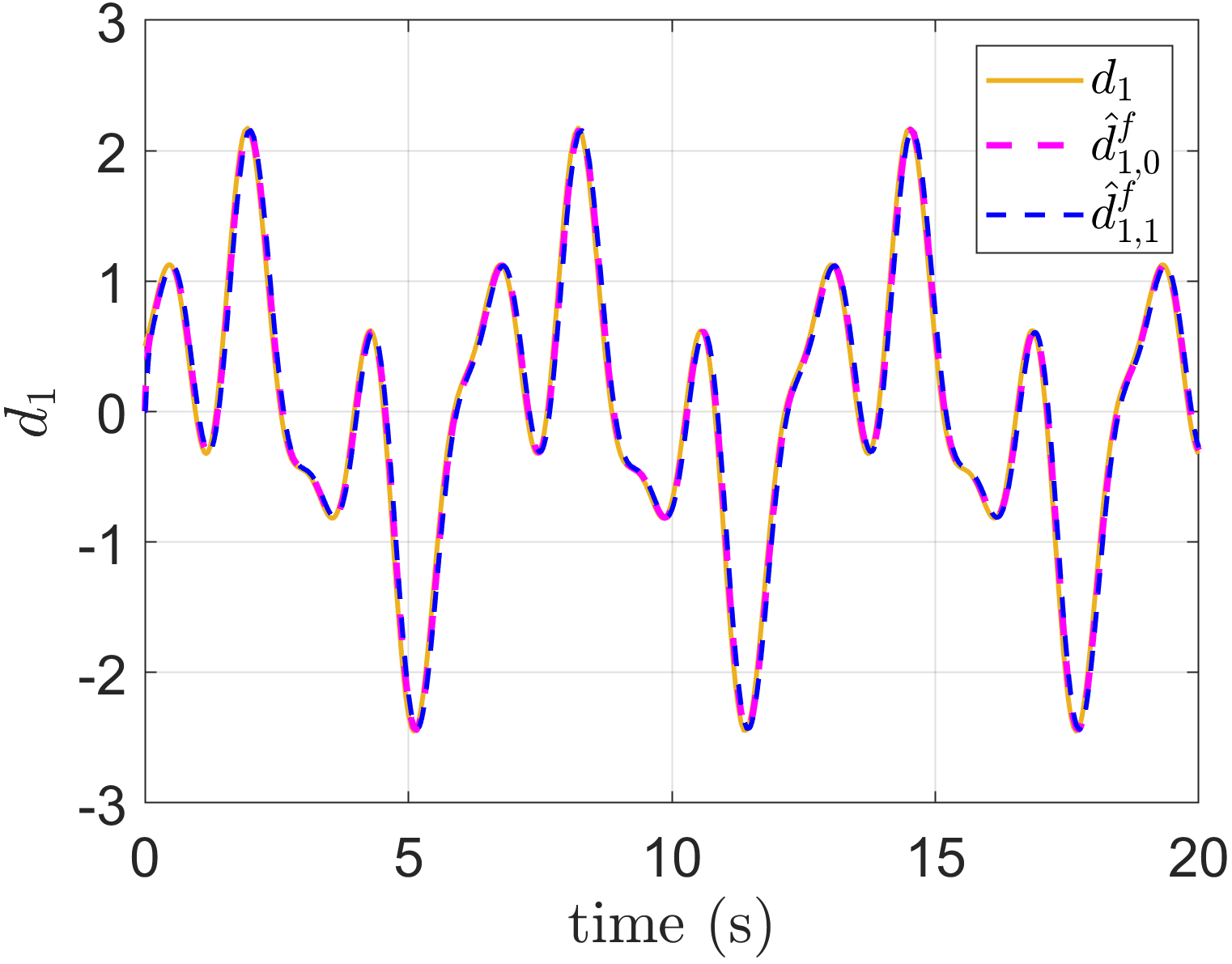}
\includegraphics[width=0.23\textwidth]{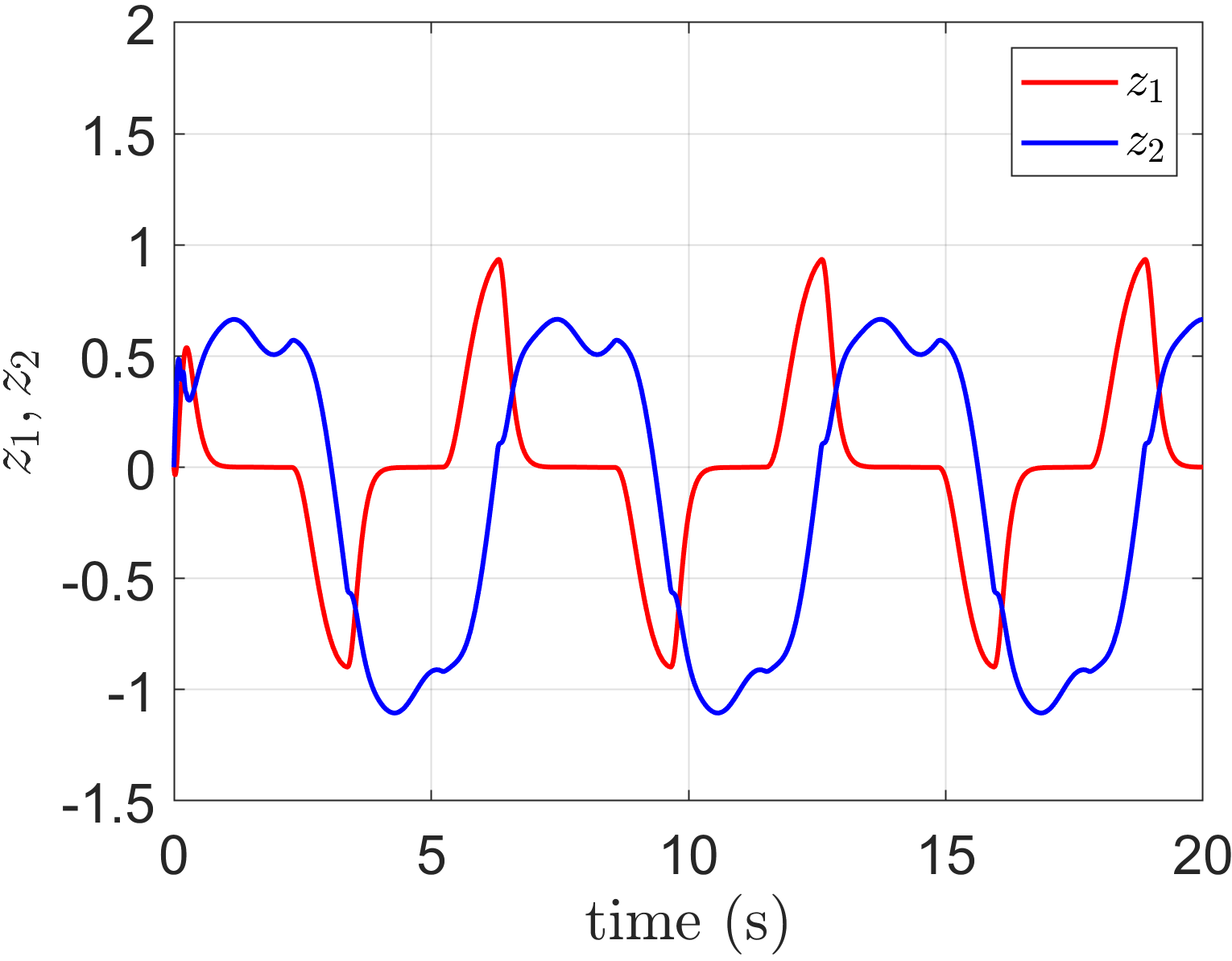}
  \caption{Simulation results of  Example \ref{example:mismatch}. All three methods can ensure safety but the DOB-PCBF method has the best tracking performance and the smoothest control profile.
  }
  \label{fig:electromechanical}
\end{figure}

\end{example}

\section{Conclusion}
\label{sec:conclusion}
A PCBF safe control design scheme that integrates Lyapunov-based and CBF-based control approaches is proposed for strict-feedback systems with potentially unknown dynamics. Moreover, a DOB-PCBF-based controller is presented for systems with mismatched disturbances. Simulation results demonstrate the effectiveness of the proposed method. Future work includes conducting experimental studies and taking input constraints into account.

\appendix

\subsection{Proof of Theorem \ref{theorem:proxy}}\label{proofthm1}
The proof consists of four steps. We will first show $b_i$ in \eqref{hi} can be expressed as the sum of three functions, and then the non-emptiness of $K_{BF}$. After that we will show $b_m\geq 0$, and $b_i\geq 0 \implies b_{i-1}\geq 0$ holds for $i\in[m]$. Finally, we can see $b_0\geq 0 \Leftrightarrow h\geq 0,\forall t\geq 0$ from the definition of $b_0$ and the function $\chi$ under \eqref{mi}.

\emph{(i) Expression of $b_i$.} 
We claim that $b_i$ in \eqref{hi} can be expressed as the sum of three functions. \\
\textbf{Claim.} The function $b_i$ defined in \eqref{hi} can be expressed in the following form for any $i\in[m]$:
\begin{equation}
    b_i = s_i(x,\bar\mu_i,\bar y_i)+l_i(y_0)+T_i(t)\label{expressionbi}
\end{equation}
where $s_i=\sum_{k=1}^{n_i}\varphi_{i}^k(x,\bar\mu_i)\bar y_i^{\alpha_i^k}$ is  a polynomial function of $\bar y_i$ for fixed $x,\bar\mu_i$,  $l_i=\prod_{j=1}^{i}\la_j y_0$ is a linear function of $y_0$, and $T_i=-\frac{\be_i}{2}\rho^2-\sum_{j=2}^{i}\frac{\be_{j-1}}{2}\left(\frac{\di}{\di t}+\la_{j} \right)\circ \cdots\circ \left(\frac{\di}{\di t}+\la_{i}\right)\circ \rho(t)^2$ is a time-varying function. In the expression of $s_i$, $n_i$ is a non-negative integer, $\varphi_i^k$ is a  function whose form can be uniquely determined, $\bar y_i^{\alpha_i^k}=y_1^{\alpha_{i1}^k}y_2^{\alpha_{i2}^k}\dots y_i^{\alpha_{ii}^k}$ where $\bar y_i=[y_1\ y_2\ \cdots\ y_i]^\top$ and  $\alpha_i^k=[\alpha_{i1}^k\ \alpha_{i2}^k\ \dots\ \alpha_{ii}^k]^\top$ is a vector of non-negative integers satisfying $\|\alpha_i^k\|_1\geq 1$, which implies that $\varphi_{i}^k(x,\bar\mu_i)\bar y_i^{\alpha_i^k}$ can not be a constant for fixed $x$ and $\bar\mu_i$. 

We will prove the claim using mathematical induction. 
For $i=1$, from \eqref{hi} we can rewrite $b_1$ as follows:
\begin{equation}
    \!\!b_1\!=\! \frac{1}{\xi} (L_{f_0}h\!+\!L_{g_0}h\mu_1)y_1\!-\!\frac{\|L_{g_0}h\|^2}{2\be_1\xi^2}y_1^2\!+\!\la_1y_0-\frac{\be_1}{2}\rho^2,
        \label{expressionb1}
\end{equation}
which implies that \eqref{expressionbi} holds when $i=1$. 

Now assume that  \eqref{expressionbi} holds for $i=d\geq1$. We only have to prove that \eqref{expressionbi} holds for $i=d+1$. 
From \eqref{mi}, we have 
$\mathcal{M}_{d+1} = \left(\frac{1}{\xi}\frac{\pa h}{\pa x}\prod_{j=1}^d\la_j\right)y_1+\frac{1}{\xi}\frac{\pa h}{\pa x}\sum_{j=1}^d\!\sum_{k=1}^{n_d}\!\varphi_d^k\frac{\pa(\bar y_d^{\alpha_d^k})}{\pa y_j}y_{j+1}
+\sum_{k=1}^{n_d}\frac{\pa \varphi_{d}^k}{\pa x}\bar y_d^{\alpha_d^k}.$
It is clear that every term in any entry of $\mathcal{M}_{d+1}$ (note that $\mathcal{M}_{d+1}$ is a vector) can be expressed in the form of $\varphi_{{d+1}}^k(x,\bar\mu_{d+1})\bar y_{d+1}^{\alpha_{d+1}^k}$ and it can not be a constant for fixed $x$, $\bar\mu_{d+1}$.
It is also easy to verify that each term in $\sum_{j=1}^{d}\frac{\pa b_{d}}{\pa\mu_j}\mu_{j+1}$ can be expressed in the form of $\varphi_{{d+1}}^k(x,\bar\mu_{d+1})\bar y_{d+1}^{\alpha_{d+1}^k}$ because $\frac{\pa b_{d}}{\pa\mu_j}\mu_{j+1}=\sum_{k=1}^{n_{d}}\left(\frac{\pa \varphi_{d}^k}{\pa\mu_j}\mu_{j+1}\right)\bar y_{d}^{\alpha_{d}^k}$. Hence, $\mathcal{M}_{d+1}(f_0+g_0\mu_1)-\frac{\|\mathcal{M}_{d+1}g_0\|^2}{2\be_{d+1}}+\sum_{j=1}^{d}\frac{\pa b_{d}}{\pa\mu_j}\mu_{j+1}$, which is included as part of $b_{d+1}$, can be expressed in the form of $s_{d+1}$. 

In addition, since $\frac{\pa b_d}{\pa t}+\la_{d+1}b_d-\frac{\be_{d+1}}{2}\rho^2$, which are the rest terms in $b_{d+1}$, are equal to $\sum_{k=1}^{n_d}(\la_{d+1}\varphi_{k}^d)\bar y_d^{\alpha_{d}^k}+\prod_{j=1}^{d+1}\la_j y_0-\frac{\be_{d+1}}{2}\rho^2-\sum_{j=2}^{d+1} \frac{\be_{j-1}}{2} \big( \frac{\di}{\di t}+\la_{j}  \big) \circ \cdots\circ \big(\frac{\di}{\di t}+\la_{d+1} \big)\circ\rho^2$,  one can express the first summation term in the form of $s_{d+1}$, consider the second term $\prod_{j=1}^{d+1}\la_j y_0$  as $l_{d+1}$ and the sum of the rest two terms as $T_{d+1}$. Hence, $b_{d+1}$ can be expressed as \eqref{expressionbi}, which completes the proof of the claim.

\emph{(ii) Non-emptiness of $K_{BF}$.}  We will show the non-emptiness of $K_{BF}$ by proving that $\psi_1=0\implies \psi_0\geq 0$ holds for any $x\in\C$ and $\mu_1,\cdots,\mu_m\in \R^{p_1}$. First we claim that $\frac{\pa b_i}{\pa\mu_i}=\frac{y_1}{\xi}L_{g_0}h$ for $i\in[m]$. For $i=1$, this claim is clear from $b_1$ given in \eqref{expressionb1}. Assume that the claim is true for $i=d$ where $d\geq 1$, then, because the only term in $b_{d+1}$ including $\mu_{d+1}$ is $\frac{\pa b_{d}}{\pa \mu_{d}}\mu_{d+1}$, implying that $\frac{\pa b_{d+1}}{\pa \mu_{d+1}}=\frac{\pa b_{d}}{\pa \mu_{d}}=\frac{y_1}{\xi}L_{g_0}h$ and thus the claim is true for $i=d+1$. Hence, the claim is proved.

Next, invoking \eqref{psi01:1}, one can see $\psi_1=\frac{y_1}{\xi}L_{g_0}h$. Now we assume $\psi_1=0$, which implies $y_1=0$ or $L_{g_0}h=0$. If $y_1=0$, one can see that $h\geq \xi$ from the definition of $\chi$. Meanwhile, from Condition (i) one can see that $L_{g_0}h=0$ indicates $h\geq \xi$ for any $x\in\C$. Note that $h\geq \xi$ implies $y_0=1$ and $y_1=y_2=\cdots=y_{m+1}=0$, according to their definitions shown below \eqref{hi}. Hence, from \eqref{expressionbi} one can conclude  $b_m=\prod_{j=1}^m\la_j -\frac{\be_m}{2}\rho^2-\sum_{j=2}^{m}\frac{\be_{j-1}}{2}\left(\frac{\di}{\di t}+\la_{j} \right)\circ \cdots\circ \left(\frac{\di}{\di t}+\la_{m}\right)\circ \rho^2$ because $s_m$ is a polynomial of $\bar y_m$ for fixed $x$ and $\bar\mu_m$ (with no constant terms), meaning that it equals $0$ as $y_i=0$, $i\in[m]$. Moreover, invoking \eqref{expressionbi}, one can readily verify that $\frac{\pa b_m}{\pa x}=\frac{\pa s_m}{\pa x}=0$ and $\sum_{j=0}^m\frac{\pa b_m}{\pa y_j}\frac{\pa h}{\pa x}\frac{y_{j+1}}{\xi}=0$, implying that $\mathcal{M}_{m+1}=0$. Similarly, it can be shown $\sum_{j=1}^{m-1}\frac{\pa b_m}{\pa \mu_j}\mu_{j+1}=0$. Thus, according to \eqref{psi01:0} and \eqref{expressionbi}, one can see $\psi_0= \frac{\pa b_m}{\pa t}+\la_{m+1}b_m=\frac{\pa T_m}{\pa t}+ \la_{m+1}b_m=\prod_{j=1}^{m+1}\la_j-\sum_{j=2}^{m+1}\frac{\be_{j-1}}{2}\left(\frac{\di}{\di t}+\la_{j} \right)\circ \cdots\circ \left(\frac{\di}{\di t}+\la_{m+1} \right)\circ \rho^2.$
From Condition (ii), $\psi_0\geq 0$, meaning that $K_{BF}\neq\emptyset$. 

\emph{(iii) $\nu\in K_{BF}\implies b_m\geq 0$.} 
One can see 
$\dot b_{m}+\la_{m+1} b_{m}
=  \frac{\pa b_{m}}{\pa x}\dot x+\sum_{j=1}^{m-1}\frac{\pa b_{m}}{\pa \mu_j}\mu_{j+1}+\frac{\pa b_{m}}{\pa t}+\frac{\pa b_{m}}{\pa \mu_m} \nu+\sum_{j=0}^{m}\frac{\pa b_{m}}{\pa y_j}\frac{y_{j+1}}{\xi}\frac{\pa h}{\pa x}\dot x 
+\la_{m+1} b_{m} 
\geq\mathcal{M}_{m+1}(f_0+g_0\mu_1)+\sum_{j=1}^{m-1}\frac{\pa b_{m}}{\pa \mu_j}\mu_{j+1}+\frac{\pa b_{m}}{\pa t}+\frac{\pa b_{m}}{\pa \mu_m} \nu+\la_{m+1} b_m-\|\mathcal{M}_{m+1} g_0\|\rho
= \psi_0+\psi_1 \nu.$
Hence, selecting $\nu\in K_{BF}$ implies $\dot b_{m}\!+\!\la_{m+1} b_{m}\!\geq\! 0$, and thus $b_{m}\geq 0$ as Condition (iii) holds.

\emph{(iv) $b_i\geq 0 \Rightarrow b_{i-1}\geq 0$ for any $i\in[m]$.} 
Note that
\begin{IEEEeqnarray}{rCl}
\dot b_{i-1}\!&=&\! \frac{\pa b_{i\!-\!1}}{\pa x}\dot x\!+\!\!\sum_{j=1}^{i-1}\!\frac{\pa b_{i\!-\!1}}{\pa \mu_j}\mu_{j+1}\!+\!\!\frac{\pa b_{i-1}}{\pa t} \!+\!\!\sum_{j=0}^{i-1}\!\frac{\pa b_{i\!-\!1}}{\pa y_j}\frac{y_{j\!+\!1}}{\xi}\frac{\pa h}{\pa x}\dot x\nonumber\\
&\geq& \sum_{j=1}^{i-1}\frac{\pa b_{i-1}}{\pa \mu_j}\mu_{j+1}+\frac{\pa b_{i-1}}{\pa t}- \frac{\|\mathcal{M}_i g_0\|^2}{2\beta_i}-\frac{\beta_i}{2}\rho^2\nonumber\\
&& +\mathcal{M}_i(f_0+g_0\mu_1) = b_i-\la_ib_{i-1}.
\end{IEEEeqnarray}
Thus, $b_i\geq 0$ indicates $\dot b_{i-1}+\la_i b_{i-1}\geq 0$, which implies $b_{i-1}\geq 0$ as $b_{i-1}(\bar\mu_{i-1}(0),\bar y_{i-1}(0),y_0(0),x(0),0)>0$. Hence, $b_0\geq 0$, which indicates that  $h(x(t))\geq 0,\forall t\geq 0$.

\subsection{Proof of Corollary \ref{corollary:tracking}}\label{proofcorollarytracking}
This proof follows the backstepping technique \cite{KKK95}. We define $V_0=\fot \ep_0^2$ and $V_i=V_0+\fot \sum_{j=1}^i\ep_j^\top \ep_j$ where $i\in[m]$ and consider the worst-case of $e$ in virtual control design.

Clearly, 
$\dot V_0=\ep_0^\top(f_0+g_0(\alpha_1+\ep_1+e)-\dot x_d)=-k_0\|\ep_0\|^2+\ep_0^\top g_0\ep_1+\ep_0^\top g_0 e-\frac{\|g_0^\top\ep_0\|^2}{2c_0}\leq-k_0\|\ep_0\|^2+\ep_0^\top g_0\ep_1+\|\ep_0^\top g_0\|\rho-\frac{\|\ep_0^\top g_0\|^2}{2c_0}\leq -k_0\|\ep_0\|^2+\ep_0^\top g_0\ep_1+\frac{c_0}{2}\bar\rho^2$, where $\bar \rho = \sup_{t\geq 0}\rho(t)$. We claim the following inequality holds 
for any $i\in[m-1]$:
\begin{equation}\label{dotvi}
    \dot V_i\leq -\sum_{j=0}^ik_j\|\ep_j\|^2+\ep_i^\top\ep_{i+1}+\frac{\bar\rho^2}{2}\sum_{j=0}^i c_j.
\end{equation}
Indeed, \eqref{dotvi} holds for $i=1$ as $\dot V_1= \dot V_0+\ep_1^\top\big(\ep_2-g_0^\top\ep_0-k_1\ep_1-\frac{\ep_1}{2c_1}\big\|\frac{\pa\alpha_1}{\pa x}g_0 \big\|^2-\frac{\pa\alpha_1}{\pa x}g_0e\big)\leq -k_0\|\ep_0\|^2-k_1\|\ep_1\|^2+\ep_1^\top\ep_2+\frac{c_0+c_1}{2}\bar\rho^2$. Now assume that \eqref{dotvi} holds for $i=k-1$ where $k\geq 2$. Since 
$\dot V_{k}=\dot V_{k-1}+\ep_k^\top\big(\ep_{k+1}-\ep_{k-1}-\frac{\ep_k}{2c_k}\big\|\frac{\pa \alpha_k}{\pa x}g_0\big\|^2
-k_k\ep_k-\frac{\pa\alpha_k}{\pa x}g_0e\big)\leq-\sum_{j=0}^{k} k_j\|\ep_j\|^2+\ep_k^\top\ep_{k+1}+\frac{\bar\rho^2}{2}\sum_{j=0}^{k} c_j$,  \eqref{dotvi} holds for $i=k$. Hence, by mathematical induction,   \eqref{dotvi} holds for all $i\in[m-1]$.

Note that 
\begin{IEEEeqnarray}{rCl}
    \dot V_m
    \!&\stackrel{\eqref{dotvi}}{\leq} &\! -\!\!\sum_{j=0}^{m-1}\! k_j\|\ep_j\|^2\!+\!\ep_{m-1}^\top\ep_m\!+\!\frac{\bar\rho^2}{2}\!\!\sum_{j=0}^{m-1}\! c_j \!+\!\ep_m(\nu_d\!-\!\dot\alpha_{m})\nonumber\\
    &\leq&\!-\!\! \sum_{j=0}^{m} \!k_j\|\ep_j\|^2\!+\!\frac{\bar\rho^2}{2}\!\sum_{j=0}^{m}\! c_j\!\leq-\chi V_m\!+\!\frac{\bar\rho^2}{2}\!\!\sum_{j=0}^{m}\! c_j,
\end{IEEEeqnarray}
where $\chi=2\min\{k_0,k_1,\cdots,k_m\}$. Using the standard Lyapunov argument, one can see that $\ep_0$ is globally UUB.

\subsection{Proof of Lemma \ref{corollary:filter}}
\label{proof:corollary3}

Define $ E_i^f=[e^{f\top}_{i,1}\ e^{f\top}_{i,2}\ \cdots\ e^{f\top}_{i,n-i}\ e_{d,i}^{\top}]^\top$ 
where $e_{d,i}$ is defined after \eqref{dob} and $e^f_{i,j}=\hat d^f_{i,j}-\hat d^f_{i,j-1}$
for $j\in[n-i]$ and $i\in [n-1]$. We will demonstrate that $\delta_i$ is globally UUB by establishing the boundedness of $E_i^f$.

From \eqref{lowpassfilter} we have 
\begin{equation}
    \dot{E}_i^f=A_iE_i^f+B_i\dot{d}_i
\end{equation}
where $B_i=[\bm{0}\ \cdots\ \bm{0}\ -I_{p_i}]^\top$ and
\begin{equation*}
   A= \begin{bmatrix}
        - T_{i,1}I_{p_i}&\hp\bm{0}&\hp\cdots&\hp\bm{0}&\hp\alpha_iI_{p_i}\\
         T_{i,1}I_{p_i}&-T_{i,2}I_{p_i}&\hp\cdots&\hp\bm{0}&\hp\bm{0}\\
    
        \vdots&\hp\vdots&\hp\ddots&\hp\vdots&\hp\vdots\\
        \hp\bm{0}&\hp\bm{0}& T_{i,n-i-1}I_{p_i}&-T_{i,n-i}I_{p_i}&\hp\bm{0}\\
        \bm{0}&\hp\cdots&\hp\bm{0}&\hp\bm{0}&\hp-\alpha_iI_{p_i}
    \end{bmatrix} .
\end{equation*}
It is easy to check that all eigenvalues of $A_i$ are negative, so for any $\ga_i>0$, there exists a positive definite matrix $P_i$ satisfying $A_i^\top P_i+P_iA_i=-\ga_i I_{p_i(n-i+1)}$. Define a candidate Lyapunov function as 
\begin{equation}
    V_i^f=E_i^{f\top} P_i E_i^f\label{vif}
\end{equation}
whose derivative satisfies
\begin{IEEEeqnarray}{rCl}
    \dot V_i^f\!&=&\! E_i^{f\top} (A_i^\top P_i\!+\!P_i A_i) E_i^f\!+\!\dot d_i^\top B_i^\top P_iE_i^f\!+\!E_i^{f\top}P_iB_i\dot d_i\nonumber\\
    &\leq& -\ga_i\|E_i^f\|^2+2\sqrt{p_i}\om_i\|P_i\|\|E_i^f\|\nonumber\\
    &\leq& -\ga_i^f\| E_i^f\|^2+\om^f_i\label{filtervfi}
\end{IEEEeqnarray}
where $0<\ta_i<\ga_i$, $\ga_i^f=\ga_i-\ta_i>0$, and $\om_i^f=\frac{p_i\om_i^2\|P_i\|^2}{\ta_i}$. From \eqref{filtervfi} one can conclude that 
$E_i^f$ is globally UUB, which indicates that $\delta_i$ is also globally UUB since
\begin{IEEEeqnarray}{rCl}
    \|\delta_i\|\leq \|e_{d,i}\|+\sum_{j=1}^{n-i} \|e_{i,j}^f\|\leq \sqrt{n-i+1} \|E_i^f\|.\label{deltai}
\end{IEEEeqnarray}

\subsection{Proof of Theorem \ref{theorem:systemmismatchblf}}
\label{proof:theorem2}
Note that for $ i\in[n-1]$, $\tau_i$ is a function of $x$,$\bar z_i$,$\bar\mu_{i+1}$,$t$,\\$\hat d_{j,n-j}^f$,$\dots$, $\hat d_{j,n-i}^f$ $(j\in[i])$. 
Define candidate BLFs $V_i (i=2,\cdots,n)$ as
\begin{equation}
\!\!\!\!   V_i(e,\bar \ep_i,\bar E_i^f)\! =\! V_{i-1}(e,\bar\ep_{i-1},\bar E_{i-1}^f)+\fot \|\ep_i\|^2\!+\!\sum_{j=1}^i\! V_j^f,\label{appendix:vi}
\end{equation}
where $V_1(e,E_1^f)=\fot\log \left(\frac{\rho^2}{\rho^2-\|e\|^2}\right)+V_1^f$, $V_i^f$ $(i\in[n-1])$ are defined in \eqref{vif}, $V_n^f=V_n^d$, and $V_n^d$ is defined after \eqref{dob}. The control design follows the BLF backstepping technique \cite{tee2009barrier}, and the disturbances are compensated by their estimates presented in \eqref{dob} and \eqref{lowpassfilter}.

Clearly $\dot V_1$ in
the open set $\mathcal{Z}(t)=\{e \in\R: \|e\|<\rho\}$ satisfies $\dot V_1\stackrel{\eqref{filtervfi}}{\leq}\frac{e^\top}{\rho^2-\|e\|^2} \big(f_1+g_1\tau_1+g_1\ep_2+d_1-\mu_2-\frac{\dot\rho}{\rho}e\big)-\ga_1^f\| E_1^f\|^2+\om_1^f\stackrel{\eqref{systemmismatchcontroller:1}}{=}\frac{e^\top}{\rho^2-\|e\|^2}(-k_1 e+g_1\ep_2-\delta_1)-\sigma_1\|E_1^f\|^2-\frac{n\|e\|^2}{4(\ga_1^f-\sigma_1)(\rho^2-\|e\|^2)^2}-(\ga_1^f-\sigma_1)\|E_1^f\|^2+\om_1^f
\stackrel{\eqref{deltai}}{\leq} -\frac{k_1\|e\|^2}{\rho^2-\|e\|^2}+\frac{e^\top g_1\ep_2}{\rho^2-\|e\|^2}+\frac{\sqrt{n}\|e\|\|E_1^f\|}{\rho^2-\|e\|^2}-\sigma_1\|E_1^f\|^2+\om_1^f -\frac{n\|e\|^2}{4(\ga_1^f-\sigma_1)(\rho^2-\|e\|^2)^2}-(\ga_1^f-\sigma_1)\| E_1^f\|^2\leq-\frac{k_1e^2}{\rho^2-\|e\|^2}-\sigma_1\|E_1^f\|^2+\frac{e^\top g_1\ep_2}{\rho^2-\|e\|^2}+\om_1^f$. Then, we will show  for any $i=2,\cdots,n-1$,
\begin{IEEEeqnarray}{rCl}
    \dot V_i\!&\leq&\!\frac{-k_1\|e\|^2}{\rho^2\!-\!\|e\|^2}\!-\!\!\sum_{j=2}^i\! k_j\|\ep_j\|^2\!\!-\!\!\sum_{j=1}^i\!\sigma_j\|E_j^f\|^2\!\!+\!\!\sum_{j=1}^i\!(i\!-\!j\!+\!1)\om_j^f\nonumber\\
    &&+\ep_i^\top g_i\ep_{i+1}\label{dotviconclusion}
\end{IEEEeqnarray} 
One can verify that $\dot V_2=\dot V_1+\ep_2^\top(\dot z_2-\dot \tau_1)+\dot V_1^f+\dot V_2^f=\dot V_1+\ep_2^\top\big(f_2+g_2\tau_2+g_2\ep_3+d_2-\frac{\pa \tau_1}{\pa x}\dot x-\frac{\pa \tau_1}{\pa z_1}\dot z_1-\sum_{j=1}^2\frac{\pa \tau_1}{\pa \mu_j}\mu_{j+1}-\frac{\pa \tau_1}{\pa \hat d_{1,n-1}^f}\dot{\hat{d}}_{1,n-1}^f-\frac{\pa \tau_1}{\pa t}\big)+\dot V_1^f+\dot V_2^f\stackrel{\eqref{systemmismatchcontroller:i},\eqref{filtervfi}}{\leq}-\frac{k_1\|e\|^2}{\rho^2-\|e\|^2}-k_2\|\ep_2\|^2-\sigma_1\|E_1^f\|^2-\sigma_2\|E_2^f\|^2+2\om_1^f+\om_2^f-(\ga_2^f-\sigma_2)\|E_2^f\|^2+\|\ep_2\|\sqrt{n-1}\|E_2^f\|-\frac{(n-1)\|\ep_2\|^2}{4(\ga_2^f-\sigma_2)}-\ga_1^f\|E_1^f\|+\|\ep_2\|\big\|\frac{\pa\tau_1}{\pa z_1}\big\|\sqrt{n}\|E_2^f\|-\frac{n\ep_2^2}{4\ga_1^f}\big\|\frac{\pa\tau_1}{\pa z_1}\big\|^2+\ep_2^\top g_2\ep_3
\leq -\frac{k_1\|e\|^2}{\rho^2-\|e\|^2}-k_2\|\ep_2\|^2-\sigma_1\|E_1^f\|^2-\sigma_2\|E_2^f\|^2+2\om_1^f+\om_2^f+\ep_2^\top g_2\ep_3$, so \eqref{dotviconclusion} holds for $i=2$. Assume \eqref{dotviconclusion} holds for $i=k-1$. Since $\dot V_k=\dot V_{k-1}+\sum_{j=1}^k\dot V_j^f+\ep_k^\top(\dot z_k-\dot\tau_{k-1})
\stackrel{\eqref{filtervfi},\eqref{dotviconclusion}}{\leq}-\frac{k_1\|e\|^2}{\rho^2-\|e\|^2}-\sum_{j=2}^{k-1}k_j\|e_j\|^2-\sum_{j=1}^{k-1}\sigma_j\|E_j^f\|^2+\sum_{j=1}^k(k-j+1)\om_j^f+\ep_{k-1}^\top g_{k-1}\ep_k-\sum_{j=1}^k\ga_j^f\|E_j^f\|^2+\ep_k^\top(f_k+g_k\tau_k+g_k\ep_{k+1}+d_k-\dot\tau_{k-1})
\stackrel{\eqref{systemmismatchcontroller:i}}{\leq}-\frac{k_1\|e\|^2}{\rho^2-\|e\|^2}-\sum_{j=2}^{k}k_j\|e_j\|^2-\sum_{j=1}^{k}\sigma_j\|E_j^f\|^2+\sum_{j=1}^k(k-j+1)\om_j^f-\sum_{j=1}^{k-1}\ga_j^f\|E_j^f\|^2-(\ga_k^f-\sigma_k)\|E_k^f\|^2+\ep_k^\top\big[g_k\ep_{k+1}-\delta_k+\sum_{j=1}^{k-1}\frac{\pa\tau_{k-1}}{\pa z_j}\delta_j-\sum_{j=1}^{k-1}\frac{(n-j+1)\ep_k}{4\ga_j^f}\big\|\frac{\pa\tau_{k-1}}{\pa z_j}\big\|^2-\frac{(n-k+1)\ep_k}{4(\ga_k^f-\sigma_k)}\big]
\stackrel{\eqref{deltai}}{\leq} -\frac{k_1\|e\|^2}{\rho^2-\|e\|^2}-\sum_{j=2}^{k}k_j\|e_j\|^2-\sum_{j=1}^{k}\sigma_j\|E_j^f\|^2+\sum_{j=1}^k(k-j+1)\om_j^f-\sum_{j=1}^{k-1}\ga_j^f\|E_j^f\|^2-(\ga_k^f-\sigma_k)\|E_k^f\|^2+\ep_k^\top g_k\ep_{k+1}+\|\ep_k\|\sqrt{n-k+1}\|E_k^f\|-\frac{(n-k+1)\|\ep_k\|^2}{4(\ga_k^f-\sigma_k)}+\sum_{j=1}^{k-1}\big\|\frac{\pa\tau_{k-1}}{\pa z_j}\big\|\times\\\|\ep_k\|\sqrt{n-j+1}\|E_j^f\|-\sum_{j=1}^{k-1}\frac{(n-j+1)\|\ep_k\|^2}{4\ga_j^f}\big\|\frac{\pa\tau_{k-1}}{\pa z_j}\big\|^2
\leq -\frac{k_1\|e\|^2}{\rho^2-\|e\|^2}-\sum_{j=2}^k k_j\|\ep_j\|^2-\sum_{j=1}^k\sigma_j\|E_j^f\|^2+\sum_{j=1}^k(k-j+1)\om_j^f+\ep_k^\top g_k\ep_{k+1}$, \eqref{dotviconclusion} holds for $i=k$. By induction, \eqref{dotviconclusion} holds  for $i=2,\cdots,n-1$. From \eqref{appendix:vi} one can see 
\begin{equation}
    \!\!\!\!V_n\!=\!\fot\!\log\! \left(\!\frac{\rho^2}{\rho^2\!-\!\|e\|^2}\!\!\right)\!+\fot\!\sum_{j=2}^n\!\|\ep_j\|^2\!+\!\!\sum_{j=1}^n\!(n-j+1)V_j^f.
\end{equation}
Meanwhile, it is easy to verify
\begin{IEEEeqnarray}{rCl}
    \dot V_n\!&\leq&\!\frac{-k_1\|e\|^2}{\rho^2\!-\!\|e\|^2}\!-\!\!\sum_{j=2}^n\! k_j\|\ep_j\|^2\!\!-\!\!\!\sum_{j=1}^n\!\sigma_j\|E_j^f\|^2\!\!-\!\!\!\sum_{j=1}^n\!(n\!-\!j\!+\!1)\om_j^f\nonumber\\
    \!&\leq&\!-k_1\log \left(\frac{\rho^2}{\rho^2-\|e\|^2}\right)-\sum_{j=2}^n k_j\|\ep_j\|^2-\sum_{j=1}^n\sigma_j\|E_j^f\|^2\nonumber\\
    &&+\sum_{j=1}^n(n-j+1)\om_j^f,
\end{IEEEeqnarray}
where $\om_n^f=\om_n^d$, $E_n^f=e_{d,n}$, $\om_n^d$ and $e_{d,n}$ are defined after \eqref{dob}, the first inequality follows a similar procedure above, and the second inequality is derived from $\log\big(\frac{\rho^2}{\rho^2-\|e\|^2}\big)\leq \frac{\|e\|^2}{\rho^2-\|e\|^2}$, which holds in $\mathcal{Z}(t)$ \cite{tee2009barrier}. Hence, $\dot V_n\leq -\zeta V_n+\sum_{j=n}^i(i-j+1)\om_j^f$, where $\zeta = \min \big\{2k_1,\cdots,2k_n,\min_{j\in[n-1]}\!\big\{\!\frac{\sigma_j}{(n-j+1)\la_{\rm min}(P_j)}\!\big\},2\sigma_n\big\}$. Thus, $V_n$ is bounded, implying  $\|e\|\leq\rho$ for any $t\geq 0$, according to \cite[Lemma 1]{tee2009barrier}.

\bibliographystyle{abbrv}
\bibliography{proxycbf.bib}

\begin{thebibliography}{10}

\bibitem{ames2016control}
A.~D. Ames, X.~Xu, J.~W. Grizzle, and P.~Tabuada.
\newblock Control barrier function based quadratic programs for safety critical systems.
\newblock {\em IEEE Transactions on Automatic Control}, 62(8):3861--3876, 2016.

\bibitem{bechlioulis2008robust}
C.~P. Bechlioulis and G.~A. Rovithakis.
\newblock Robust adaptive control of feedback linearizable {MIMO} nonlinear systems with prescribed performance.
\newblock {\em IEEE Transactions on Automatic Control}, 53(9):2090--2099, 2008.

\bibitem{bechlioulis2014low}
C.~P. Bechlioulis and G.~A. Rovithakis.
\newblock A low-complexity global approximation-free control scheme with prescribed performance for unknown pure feedback systems.
\newblock {\em Automatica}, 50(4):1217--1226, 2014.

\bibitem{breeden2021high}
J.~Breeden and D.~Panagou.
\newblock High relative degree control barrier functions under input constraints.
\newblock In {\em IEEE Conference on Decision and Control}, pages 6119--6124, 2021.

\bibitem{chen2015disturbance}
W.-H. Chen, J.~Yang, L.~Guo, and S.~Li.
\newblock Disturbance-observer-based control and related methods -- {A}n overview.
\newblock {\em IEEE Transactions on Industrial Electronics}, 63(2):1083--1095, 2015.

\bibitem{cohen2024safety}
M.~H. Cohen, T.~G. Molnar, and A.~D. Ames.
\newblock Safety-critical control for autonomous systems: Control barrier functions via reduced-order models.
\newblock {\em Annual Reviews in Control}, 57:100947, 2024.

\bibitem{cohen2023characterizing}
M.~H. Cohen, P.~Ong, G.~Bahati, and A.~D. Ames.
\newblock Characterizing smooth safety filters via the implicit function theorem.
\newblock {\em IEEE Control Systems Letters}, 7:3890 -- 3895, 2023.

\bibitem{das2024robust}
E.~Das and J.~W. Burdick.
\newblock Robust control barrier functions using uncertainty estimation with application to mobile robots.
\newblock {\em arXiv preprint arXiv:2401.01881}, 2024.

\bibitem{du2014adaptive}
J.~Du, A.~Abraham, S.~Yu, and J.~Zhao.
\newblock Adaptive dynamic surface control with {N}ussbaum gain for course-keeping of ships.
\newblock {\em Engineering Applications of Artificial Intelligence}, 27:236--240, 2014.

\bibitem{fossen1999guidance}
T.~I. Fossen.
\newblock {\em Guidance and Control of Ocean Vehicles}.
\newblock New York, NY, USA: Wiley, 1994.

\bibitem{jankovic2018robust}
M.~Jankovic.
\newblock Robust control barrier functions for constrained stabilization of nonlinear systems.
\newblock {\em Automatica}, 96:359--367, 2018.

\bibitem{jin2018adaptive}
X.~Jin.
\newblock Adaptive fixed-time control for mimo nonlinear systems with asymmetric output constraints using universal barrier functions.
\newblock {\em IEEE Transactions on Automatic Control}, 64(7):3046--3053, 2018.

\bibitem{krstic2023inverse}
M.~Krstic.
\newblock Inverse optimal safety filters.
\newblock {\em IEEE Transactions on Automatic Control}, 69(1):16--31, 2023.

\bibitem{KKK95}
M.~Krstic, P.~V. Kokotovic, and I.~Kanellakopoulos.
\newblock {\em Nonlinear and Adaptive Control Design}.
\newblock New York, NY, USA: Wiley, 1995.

\bibitem{liu2019barrier}
W.~Liu, Y.~Wei, and G.~Duan.
\newblock Barrier {L}yapunov function-based integrated guidance and control with input saturation and state constraints.
\newblock {\em Aerospace Science and Technology}, 84:845--855, 2019.

\bibitem{liu2017barrier}
Y.-J. Liu and S.~Tong.
\newblock Barrier {L}yapunov functions for {N}ussbaum gain adaptive control of full state constrained nonlinear systems.
\newblock {\em Automatica}, 76:143--152, 2017.

\bibitem{molnar2023safety}
T.~G. Molnar and A.~D. Ames.
\newblock Safety-critical control with bounded inputs via reduced order models.
\newblock In {\em American Control Conference}, pages 1414--1421. IEEE, 2023.

\bibitem{nguyen2021robust}
Q.~Nguyen and K.~Sreenath.
\newblock Robust safety-critical control for dynamic robotics.
\newblock {\em IEEE Transactions on Automatic Control}, 67(3):1073--1088, 2021.

\bibitem{ong2019universal}
P.~Ong and J.~Cort{\'e}s.
\newblock Universal formula for smooth safe stabilization.
\newblock In {\em IEEE Conference on Decision and Control}, pages 2373--2378, 2019.

\bibitem{tan2021high}
X.~Tan, W.~S. Cortez, and D.~V. Dimarogonas.
\newblock High-order barrier functions: Robustness, safety, and performance-critical control.
\newblock {\em IEEE Transactions on Automatic Control}, 67(6):3021--3028, 2021.

\bibitem{taylor2022safe}
A.~J. Taylor, P.~Ong, T.~G. Molnar, and A.~D. Ames.
\newblock Safe backstepping with control barrier functions.
\newblock In {\em IEEE Conference on Decision and Control}, pages 5775--5782, 2022.

\bibitem{tee2009barrier}
K.~P. Tee, S.~S. Ge, and E.~H. Tay.
\newblock Barrier {L}yapunov functions for the control of output-constrained nonlinear systems.
\newblock {\em Automatica}, 45(4):918--927, 2009.

\bibitem{wang2022disturbance}
Y.~Wang and X.~Xu.
\newblock Disturbance observer-based robust control barrier functions.
\newblock In {\em American Control Conference}, pages 3681--3687, 2023.

\bibitem{wang2023safe}
Y.~Wang and X.~Xu.
\newblock Safe control of {E}uler-{L}agrange systems with limited model information.
\newblock In {\em IEEE Conference on Decision and Control}, pages 5722--5728, 2023.

\bibitem{wang2024immersion}
Y.~Wang and X.~Xu.
\newblock Immersion and invariance-based disturbance observer and its application to safe control.
\newblock {\em IEEE Transactions on Automatic Control}, 69(12):8782--8789, 2024.

\bibitem{wen2011robust}
C.~Wen, J.~Zhou, Z.~Liu, and H.~Su.
\newblock Robust adaptive control of uncertain nonlinear systems in the presence of input saturation and external disturbance.
\newblock {\em IEEE Transactions on Automatic Control}, 56(7):1672--1678, 2011.

\bibitem{xu2017correctness}
X.~Xu, J.~W. Grizzle, P.~Tabuada, and A.~D. Ames.
\newblock Correctness guarantees for the composition of lane keeping and adaptive cruise control.
\newblock {\em IEEE Transactions on Automation Science and Engineering}, 15(3):1216--1229, 2018.

\bibitem{yang2013nonlinear}
J.~Yang, S.~Li, C.~Sun, and L.~Guo.
\newblock Nonlinear-disturbance-observer-based robust flight control for airbreathing hypersonic vehicles.
\newblock {\em IEEE Transactions on Aerospace and Electronic Systems}, 49(2):1263--1275, 2013.

\bibitem{yu2018finite}
J.~Yu, P.~Shi, and L.~Zhao.
\newblock Finite-time command filtered backstepping control for a class of nonlinear systems.
\newblock {\em Automatica}, 92:173--180, 2018.

\end{thebibliography}

\end{document}